\documentclass{article}
\usepackage{amssymb}

\usepackage{graphicx}
\usepackage{amsmath}

\input{tcilatex}

\begin{document}

\title{Clifford Residues and Charge Quantization}
\author{Marcus S. Cohen \\
Department of Mathematical Sciences\\
New Mexico State University\\
Las Cruces, New Mexico\\
marcus@nmsu.edu}
\date{\today}
\maketitle

\begin{abstract}
We derive the quantization of action, particle number, and \emph{electric}
charge in a Lagrangian spin bundle over $\mathbb{M}\equiv \mathbb{M}%
_{\#}\backslash \cup D_{J}$, Penrose's conformal compactification of
Minkowsky space, with the world tubes of massive particles removed.

Our Lagrangian density, $\mathcal{L}_{g}$, is the spinor factorization of
the Maurer-Cartan $4$-form $\Omega ^{4}$; it's action, $S_{g}$, measures the
covering number of the $4$ \emph{internal} $u\left( 1\right) \times su\left(
2\right) $ phases over external spacetime $\mathbb{M}$. Under $PTC$
symmetry, $\mathcal{L}_{g}$ reduces to the second Chern form $TrK_{L}\wedge
K_{R}$ for a left $\oplus $ right chirality spin bundle. We prove a \emph{%
residue theorem} for $gl\left( 2,\mathbb{C}\right) $-valued forms, which
says that, when we ``sew in'' singular loci $D_{J}$ over which the $u\left(
1\right) \times su\left( 2\right) $ phases of the matter fields have some
extra twists compared to the $\mathbf{8}$ vacuum modes, the additional
contributions to the action, electric charge, lepton and baryon numbers are
all \emph{topologically quantized}. Because left and right chirality $2$%
-forms are \emph{chiral dual,} forms are quantized over their \emph{dual}
cycles. Thus it is the interaction $c_{2}\left( E\right) $, with a globally
nontrivial \emph{magnetic} field, that forces \emph{electric fields} to be
topologically quantized over \emph{spatial} $2$ cycles, $\int_{\mathbb{S}%
^{2}}K_{or}e^{\theta }\wedge e^{\varphi }=4\pi N$.
\end{abstract}

\section{Introduction}

Yang-Mills monopoles \cite{kron} have topologically quantized \emph{magnetic}
charges because it is the \emph{magnetic} parts, $K_{jk}e^{j}\wedge e^{k}$ ($%
j,k=1,2,3$), of their $su\left( 2\right) $-valued spin-curvature $2$ forms, $%
K=\mathbf{d}\Omega +\Omega \wedge \Omega $, that ``wrap'' integrally around 
\emph{spatial} $2$-cycles. For electric charge to be topologically
quantized, the \emph{electric} field would have to wrap integrally about its 
\emph{dual} \emph{(spatial)} cycle, 
\begin{equation*}
\int_{\mathcal{S}_{2}}K_{0r}e^{\theta }\wedge e^{\varphi }=4\pi N;
\end{equation*}
Gauss' law.

The \emph{instantons} and \emph{dyons} of Yang Mills theories, which possess
(anti-) \emph{self-dual} curvatures, also possess nonzero electric fields, $%
K_{0j}=\pm \epsilon _{j}^{k\ell }K_{k\ell }$, and electric charges that are
quantized because their magnetic charges are. However, they live in a \emph{%
Euclidean} four-space. Any analogous construction in Minkowsky space, where $%
\ast \ast =-1$, must have imaginary (anti-) self-dual curvatures $\ast K=\pm
iK$.

We exhibit a model here in which Left and Right spin curvatures are \emph{%
imaginary chiral dual}: $K_{R}=\pm i\ast K_{L}$. Localized chiral dual
solutions are dyons with half-integral units of electric and magnetic
charges. For $PT$ \emph{antisymmetric} ($PT_{A}$) solutions, the electric
semi-charges add, while the magnetic semi-charges cancel, thus binding
together the left and right chiral halves into a bispinor particle.

In the work of Van der Waerden \cite{vdw}, Sachs \cite{sachs}, Penrose \cite
{penrose}, and Keller \cite{keller1}, \cite{keller2}, it becomes clear that
geometric and Fermionic fields are the integral and half-integral sectors of
one unified spin-4 tensor field.

In a companion paper \cite{clifford} (see also \cite{capp}), we exhibited a
grand-unified Lagrangian density, 
\begin{equation}
\mathcal{L}_{g}=i\int_{\mathbb{M}}\mathbf{d}\varsigma ^{\pm }\xi _{\mp
}\wedge \chi ^{\pm }\mathbf{d}\eta _{\pm }\wedge \mathbf{d}\chi ^{\mp }\eta
_{\pm }\wedge \varsigma ^{\mp }\mathbf{d}\xi _{\pm }\text{,}  \label{1}
\end{equation}
(\emph{sum over all neutral sign combinations}) invariant under the group $%
E_{P}$ of \emph{passive }Einstein transformations; Sachs' \cite{sachs} term
for the global extension of the Poincar\'{e} group to a Friedmann universe. $%
E_{P}$ transformations connect the \emph{same} physical state in the moving
frames of different observers. In the $PTC$-symmetric \emph{geometrical
optics }$(g.o.)$ regime in $\mathbb{M}=\mathbb{M}_{\#}\backslash \cup D_{J}$%
, outside the \emph{singular loci} $D_{J}$, $\mathcal{L}_{g}$ reduces to the 
\emph{Maurer-Cartan} $4$ \emph{form}. This gives a natural topological
action 
\begin{equation}
S_{g}=i\int_{\mathbb{M}}Tr\Omega ^{L}\wedge \Omega _{L}\wedge \Omega
^{R}\wedge \Omega _{R}\equiv i\int_{\mathbb{M}}\widehat{\mathcal{L}}_{g}%
\text{,}  \label{1a}
\end{equation}
which measures the covering number of spin space over spacetime, and comes
in quantized units.

$\mathcal{L}_{g}$ of (\ref{1}) is not unique---but its action $S_{g}$ (\ref
{1a}) does have a desirable feature: The terms in $S_{g}$ decompose into
effective \emph{electroweak, strong,} and \emph{gravitational} potentials
and curvatures, together with their proper field actions \cite{clifford}. We
show here, using \emph{spin residues}---``winding numbers'' of $gl\left( 2,%
\mathbb{C}\right) $-valued forms about each codimension $J$, singularity $%
D_{J}$---that these \emph{actions} and charges are \emph{topologically
quantized.}

The \emph{singular loci} $D_{J}$ are where $J=1$, $2$, $3$, or $4$ pairs of
spin rays cross, forming \emph{caustics}. Here the $gl\left( 2,\mathbb{C}%
\right) $ \emph{phases} of $J$ chiral pairs of spinors, i.e. the local,
path-dependent exponents in the geometrical optics ($g.o.$) ansatz 
\begin{equation}
\psi \left( x\right) =e^{\frac{i}{2}\left( \theta ^{\alpha }\left( x\right)
+i\varphi ^{\alpha }\left( x\right) \right) \sigma _{\alpha }}\psi \left(
0\right) \equiv e^{\frac{i}{2}\varsigma ^{\alpha }\left( x\right) \sigma
_{\alpha }}\psi \left( 0\right) \text{,}  \label{2}
\end{equation}
cannot be defined. This happens when

\begin{enumerate}
\item  $D_{J}$ contains a \emph{zero} of $\psi \equiv \xi _{\pm }$, $\eta
_{\pm }$, $\varsigma ^{\pm }$, or $\chi ^{\pm }$;

\item  $\psi $ or $\mathbf{d}\psi $ is undefined somewhere in $D_{J}$, i.e. $%
D_{J}$ contains a \emph{singular point} of $\psi $;

\item  the phases of each field in (\ref{2}) are \emph{defined,} but $J$
pairs break away from $PTC$ conjugacy. The transformations that create these
states violate the \emph{spin isometry condition} 
\begin{equation}
\zeta ^{\pm }\xi _{\mp }=1=\chi ^{\pm }\eta _{\mp }\text{.}  \label{3}
\end{equation}

\item  $J$ of the $4$ gradients in $\mathcal{L}_{g}$ become \emph{linearly}
dependent in $D_{J}$, and so fail to span a $4$-volume element. The
remaining pairs span the $\left( 4-J\right) $-surface over which the $J$
broken out fields are quantized, as we show below.
\end{enumerate}

We call the row spinors $\varsigma ^{\mp }$ and $\chi ^{\mp }$ in (\ref{1})
the \emph{Baryonic spinors}. They must be treated as \emph{independent
variables} from the \emph{leptonic} (column) spinors $\xi _{\pm }$ and $\eta
_{\pm }$ in the variation of $\mathcal{L}_{g}$ within each singular domain $%
D_{J}$. In the companion paper \cite{clifford}, we identify codimension $J=1$%
, $2$, $3$, and $4$ \emph{topological defects} in the multi-spinor fields
with leptons, bosons, hadrons and their reaction vertices, respectively.
Inside the $D_{1}$, $\mathcal{L}_{g}$ gives Dirac equations coupling each
chiral pair of matter fields through nonlinear scatterings with the vacuum
fields, thus creating the effective masses of bispinor particles \cite
{clifford}.

However, it is not necessary to unravel the detailed structure of these core
regions to prove that they carry \emph{integral charges}---\emph{electric
charge, lepton number,} and \emph{baryon number}---and of \emph{action},
provided that the ``inner'' solutions for $\mathcal{L}_{g}$ match the
``outer'' ($g.o.$) solutions for $\widehat{\mathcal{L}}_{g}$ outside the
singular domains, i.e. in $\mathbb{M\equiv M}_{\#}\backslash \cup D_{J}$.

Below we prove a $\left( 3+1\right) $-dimensional \emph{Clifford residue
theorem} for Lie-algebra-valued forms, that says each singular domain
contributes integral units of action and charge for \emph{any} Lagrangian
density that is a natural $4$ form. The argument breaks down into four steps:

\begin{enumerate}
\item  Separate the action into outer (field) and inner (matter)
contributions, 
\begin{equation*}
S_{g}=\int_{\mathbb{M}}\widehat{\mathcal{L}}_{g}+\int_{\cup D_{J}}\mathcal{L}%
_{g}=S_{F}+S_{M}\text{.}
\end{equation*}

\item  Show that the field action for the vacuum spin bundle $\hat{\Psi}$
over the compact base space, $\mathbb{M}_{\#}\equiv \mathbb{S}^{1}\times 
\mathbb{S}^{3}$, is topologically quantized.

\item  Act on $\hat{\Psi}$ with topologically nontrivial \emph{active local }%
Einstein ($E_{A}$) transformations that may become singular in codimension-$%
J $ domains $D_{J}$.

\item  Show that the resulting \emph{field actions} and \emph{charges} are
all topologically quantized over $\mathbb{M}$.
\end{enumerate}

\section{Spin Connections and Maurer-Cartan Forms}

We briefly review how spinors factor the ``internal'' Lie-algebra $gl\left(
2,\mathbb{C}\right) $ of conformal spinors (see Appendix). The affine \emph{%
spin connection} $\Omega $ gives the spin-space increment that corresponds
to each space-time increment, and \emph{vice versa}. $\Omega $ is a $%
gl\left( 2,\mathbb{C}\right) $-valued $1$ form that enters into the
covariant derivative to assure covariance under coupled internal/external
spin transformations in any moving frame.

We specialize below to spacetime and spin frames adapted to a \emph{%
Friedmann universe; }an expanding ``$3$ brane'' $S_{3}\left( T\right) $
that, at \emph{``cosmic'' time} $T$, is approximately a hypersphere $\mathbb{%
S}^{3}\left( a\right) \subset \mathbb{R}^{4}$, with radius 
\begin{equation}
a\left( T\right) =e^{\frac{T}{a_{\#}}}a_{\#}\equiv \gamma a_{\#}\text{.}
\label{4}
\end{equation}
Here $a_{\#}$ is the equilibrium radius \cite{im}; $\gamma $ is the
conformal \emph{scale factor.}

The \emph{real} radial coordinate $T$ is not directly visible to us as
observers embedded in $S_{3}\left( T\right) $. In relativistic kinematics, $%
T $ is replaced by \emph{arctime} $x^{0}\subset \mathbb{S}^{1}$: the
arclength travelled on $\widetilde{\mathbb{S}}^{3}$ by a photon, projected
down to $\mathbb{S}^{3}\left( a_{\#}\right) $, the fiducial three-sphere of
stationery radius $a_{\#}$.

Arctime $x^{0}$ enters \cite{im} as the real part of a \emph{complex} time
coordinate $z^{0}\equiv x^{0}+iy^{0}$; cosmic time $T\equiv y^{0}$ is the
imaginary part. We do our local physics in a dilation-invariant way by
projecting down to $\mathbb{M}_{\#}\equiv \mathbb{S}^{1}\times \mathbb{S}%
^{3}\left( a_{\#}\right) $, Penrose's \cite{penrose} conformal
compactification of Minkowsky space, with canonical (Lie-algebra)
``coordinates'' $x=\left( x^{0},x^{1},x^{2},x^{3}\right) $.

$\mathbb{M}_{\#}$ is a very nice space on which to work, because it is a Lie
group: 
\begin{equation*}
\mathbb{M}_{\#}\equiv \mathbb{S}^{1}\times \mathbb{S}^{3}\sim U\left(
1\right) \times SU\left( 2\right) \text{.}
\end{equation*}
$\mathbb{S}^{3}$ has two natural representations of translation, Left ($L$)
and Right ($R$), that derive from Left or Right translation in $SU\left(
2\right) $. These are the two \emph{chiralities}.

Adding a $u\left( 1\right) $ generator $\sigma _{0}$ to each, we obtain $%
\sigma _{\alpha }\in u\left( 1\right) \times su\left( 2\right) _{L}$ and $%
\bar{\sigma}_{\alpha }\in u\left( 1\right) \times su\left( 2\right) _{R}$,
the\emph{\ left} and \emph{right} Lie algebras. These must be viewed as 
\emph{independent generators} of chiral $U\left( 1\right) \times SU\left(
2\right) $. However, note that $\bar{\sigma}_{\alpha }$ is the \emph{dual}
Lie algebra to $\sigma _{\alpha }\equiv \left( \sigma _{0},\sigma
_{1},\sigma _{2},\sigma _{3}\right) $, under the \emph{Clifford-Killing form}
for the Minkowsky metric, $\eta _{\alpha \beta }\equiv diag\left(
1,-1,-1,-1\right) $: 
\begin{equation}
\left\{ \sigma _{\alpha },\bar{\sigma}_{\beta }\right\} \equiv \sigma
_{\alpha }\bar{\sigma}_{\beta }+\sigma _{\beta }\bar{\sigma}_{\alpha }=2\eta
_{\alpha \beta }\sigma _{0}\text{.}  \label{5}
\end{equation}
\begin{equation}
\sigma ^{\alpha }=\bar{\sigma}_{\alpha }\text{;\quad }\sigma ^{\rho }\sigma
_{\rho }=-2\text{,}  \label{6}
\end{equation}
is the Lorenz-invariant form.

We may thus define the \emph{Clifford product} of ``spinorized'' tangent
vectors $a,b\in T\mathbb{M}_{\#}$, 
\begin{equation}
\begin{array}{c}
a=a^{\alpha }\sigma _{\alpha }\text{,} \\ 
\bar{b}=b^{\beta }\bar{\sigma}_{\beta }:\frac{1}{2}\left( a\bar{b}+b\bar{a}%
\right) =\eta _{\alpha \beta }a^{\alpha }b^{\beta }\sigma _{0}\equiv
a_{\beta }b^{\beta }\sigma _{0\text{.}}
\end{array}
\label{7}
\end{equation}
This is the \emph{scalar} $\sigma _{0}$ in the Lie algebra times the \emph{%
Minkowsky} product of the vectors. Note that the Clifford scalar is picked 
\emph{out} by the \emph{Trace:} 
\begin{equation}
\frac{1}{2}Tr\left( a\bar{b}\right) =a_{\beta }b^{\beta
}=a_{0}b^{0}-a_{1}b^{1}-a_{2}b^{2}-a_{3}b^{3}\text{.}  \label{8}
\end{equation}
In curved spacetime (A11), the $\eta _{\alpha \beta }$ are replaced by the
metric coefficients $g_{\alpha \beta }$.

The columns of spin frames (A5) are a basis for the fundamental $L$ and $R$
chirality spinors $\xi _{\pm }\left( x\right) $ and $\eta _{\pm }\left(
x\right) $ painted on $\mathbb{M}_{\#}$ by the \emph{spinorization maps} 
\begin{equation}
\begin{array}{c}
S:g_{\pm }\left( x\right) \equiv \exp \left( \frac{i}{2a_{\#}}x^{\alpha
}\sigma _{\alpha }^{\pm }\right) :\mathbb{S}^{1}\times \mathbb{S}%
^{3}\longrightarrow U\left( 1\right) _{\pm }\times SU\left( 2\right) _{L} \\ 
\bar{S}:\bar{g}_{\pm }\left( x\right) \equiv \exp \left( \frac{i}{2a_{\#}}%
x^{\alpha }\bar{\sigma}_{\alpha }^{\pm }\right) :\mathbb{S}^{1}\times 
\mathbb{S}^{3}\longrightarrow U\left( 1\right) _{\pm }\times SU\left(
2\right) _{R}\text{,}
\end{array}
\label{9}
\end{equation}
where $\sigma _{\alpha }^{\pm }\equiv \left( \pm \sigma _{0},\mathbf{\sigma }%
\right) $. Their infinitesimal versions are the $L$- and $R$-invariant \emph{%
Maurer-Cartan }$1$\emph{\ forms}: 
\begin{equation}
\begin{array}{c}
TS\left( x\right) \equiv g_{\pm }^{-1}dg_{\pm }\left( x\right) =\frac{i}{%
2a_{\#}}\sigma _{\alpha }^{\pm }e^{\alpha }\left( x\right) :e_{\beta }\left(
x\right) \longrightarrow \frac{i}{2a_{\#}}\sigma _{\beta }^{\pm }\left(
x\right) \\ 
T\bar{S}\left( x\right) \equiv \bar{g}_{\pm }^{-1}d\bar{g}_{\pm }\left(
x\right) =\frac{i}{2a_{\#}}\bar{\sigma}_{\alpha }^{\pm }\bar{e}^{\alpha
}\left( x\right) :\bar{e}_{\beta }\left( x\right) \longrightarrow \frac{i}{%
2a_{\#}}\bar{\sigma}_{\beta }^{\pm }\left( x\right) \text{.}
\end{array}
\label{10}
\end{equation}

The Maurer-Cartan $1$ forms give the images in the ``internal'' Lie algebras 
$u\left( 1\right) _{\pm }\times su\left( 2\right) _{L}$ and $u\left(
1\right) _{\pm }\times su\left( 2\right) _{R}$ of infinitesimal $L$ and $R$
translations on $\mathbb{M}_{\#}$; i.e. the canonical spin-space increments
that accompany a spacetime translation on $\mathbb{M}_{\#}$.

In the presence of a \emph{source,} a translation is accompanied by \emph{%
active local} spin space increments $\ell \left( x\right) $ and $r\left(
x\right) $ in the reference frame of an observer $O$. $O$ then experiences
the \emph{vector potentials} 
\begin{equation}
\begin{array}{c}
\Omega _{L}\equiv \ell ^{-1}\mathbf{d}\ell =\Omega _{L\alpha }e^{\alpha }%
\text{;} \\ 
\Omega _{L\alpha }=\ell ^{-1}\partial _{\alpha }\ell \text{;\qquad }\Omega
_{R}\equiv r^{-1}\mathbf{d}r\text{.}
\end{array}
\label{11}
\end{equation}
The Lie-algebra-valued $1$ forms, or \emph{spin connections} $\Omega _{L}$
and $\Omega _{R}$ are the \emph{Maurer-Cartan }$1$\emph{\ forms} for local $%
Gl\left( 2,\mathbb{C}\right) $ \emph{deformations} $\ell \left( x\right) $
and $r\left( x\right) $ of the canonical maps (\ref{9}) of spacetime into
spin space (see (A10) below). \emph{Regular} $g.o.$ perturbations do not
change the rank of the mapping $\psi $ of physical space to spin space.

\section{Vector Potentials from Active Local Spin Transformations}

\emph{Active local} ($E_{A}$) transformations represent both local
dilation/boost flows and local $U\left( 1\right) \times SU\left( 2\right) $
phase flows in the \emph{geometrical optics }($g.o.$) regime. $E_{A}$
transformations on the tetrads (A8), (A9) are presented as \emph{complexified%
} chiral 
\begin{equation*}
U\left( 1\right) \times SU\left( 2\right) \overset{\mathbb{C}}{%
\Longrightarrow }GL\left( 2,\mathbb{C}\right)
\end{equation*}
spin transformations on the canonical spin frames: 
\begin{equation}
\begin{array}{c}
\ell \left( z\right) =\ell \left( 0\right) L\left( z\right) \equiv \ell
\left( 0\right) \exp \frac{i}{2}\left( \theta _{L}^{\alpha }\left( z\right)
+i\varphi _{L}^{\alpha }\left( z\right) \right) \sigma _{\alpha }\equiv \ell
\left( 0\right) e^{\frac{i}{2}\varsigma _{L}^{\alpha }\left( z\right) \sigma
_{\alpha }} \\ 
\bar{r}\left( z\right) =\bar{R}\left( z\right) \bar{r}\left( 0\right) \equiv
\exp \left( \frac{i}{2}\left( \theta _{R}^{\alpha }\left( z\right) +i\varphi
_{R}^{\alpha }\left( z\right) \right) \sigma _{\alpha }\right) \bar{r}\left(
0\right) \equiv e^{\frac{i}{2}\varsigma _{R}^{\alpha }\left( z\right) \sigma
_{\alpha }}\bar{r}\left( 0\right) \text{,}
\end{array}
\label{12}
\end{equation}
where we may take $\ell \left( 0\right) =\sigma _{0}=r\left( 0\right) $.

In a spin bundle $E$ with a momentum flow $y_{\beta }\left( x\right) $, the
Cartan moving spin frames (\ref{12}) are \emph{path dependent} functions of $%
x$. The $\theta _{L}^{\alpha }\left( x\right) $ are the coefficients of the 
\emph{anti-Hermitian }($aH$) matrices $\frac{i}{2}\sigma _{\alpha }$ that
generate (local) \emph{unitary} $U\left( 1\right) \times SU\left( 2\right)
_{L}$ spin transformations. Their differentials are the \emph{electroweak
vector potentials:} 
\begin{equation*}
\frac{i}{2}\mathbf{d}\theta ^{\alpha }\sigma _{\alpha }\equiv W_{\beta
}e^{\beta }\text{.}
\end{equation*}
The $\varphi _{L}^{\alpha }\left( x\right) $ are the coefficients of the
Hermitian ($H$) generators $\frac{1}{2}\sigma _{\alpha }$ which give the
local dilation/boost flow, and whose differentials are the \emph{%
gravitational potentials,} 
\begin{equation*}
\frac{i}{2}\mathbf{d}\varphi ^{\alpha }\sigma _{\alpha }\equiv \Phi _{\beta
}e^{\beta }\text{.}
\end{equation*}
For example, the Newtonian potential $\mathbf{d}\varphi ^{0}\left( x\right) $
represents a local contraction of the spatial step corresponding to a fixed
increment in the amplitude of the spinor fields. Outside the singular loci,
we expect the phase flow to be \emph{analytic,} so the Cauchy-Riemann
equations will hold: 
\begin{equation}
\frac{\partial \zeta ^{\alpha }}{\partial \bar{z}^{\beta }}=0\Longrightarrow 
\frac{\partial \theta ^{\alpha }}{\partial x^{\beta }}=\frac{\partial
\varphi ^{\alpha }}{\partial y^{\beta }}\text{;\quad }\frac{\partial \theta
^{\alpha }}{\partial y^{\beta }}=-\frac{\partial \varphi ^{\alpha }}{%
\partial x^{\beta }}\text{.}  \label{13}
\end{equation}
There the $gl\left( 2,\mathbb{C}\right) $ phase factors $\theta ^{\alpha
}\left( z\right) $ and $\varphi ^{j}\left( z\right) $ in (\ref{12}) are
functions of $z$, the position-momentum coordinates assigned to a point in
phase space by an observer, $O$. $E$ transformations thus act \cite{im} on
the \emph{complexified} tetrads 
\begin{equation}
\begin{array}{c}
q_{\alpha }\left( z\right) \equiv \ell \left( z\right) \otimes _{\alpha }%
\bar{r}\left( z\right) \text{;} \\ 
z\equiv z^{\beta }\equiv x^{\beta }+iy^{\beta }\text{, }\beta =0,1,2,3\text{;%
} \\ 
z^{\beta }\in \mathbb{CM}\subset T^{\ast }\mathbb{M}.
\end{array}
\label{14}
\end{equation}
The $z^{\beta }$ are $4$ \emph{complex} coordinates on the Dirac \emph{phase
space}.

Just as $q\left( t\right) $ is the complex position-momentum vector for the
harmonic oscillator $q\left( t\right) =i\omega q\left( t\right) $ as a first
order system, the $q_{\alpha }$ are complex vectors in the position-momentum
frame bundle $\mathbb{CM}.$ This complex structure, along with the
antisymmetric inner product 
\begin{equation}
\left\langle \ell _{\mathbf{1}},\ell _{\mathbf{2}}\right\rangle \equiv
\left| \ell \right| =\ell _{\mathbf{1}}^{T}\epsilon \ell _{\mathbf{2}}\equiv
\ell ^{\mathbf{1}}\ell _{\mathbf{2}}\text{,}  \label{15}
\end{equation}
gives a \emph{symplectic} structure \cite{atiyah} on $T^{\ast }\mathbb{M}$.
The \emph{norm} of a spin frame is its \emph{determinant} (\ref{15}), the
area in phase space that it spans.

The canonical spin connections on $\mathbb{M}_{\#}$ are obtained for $%
y^{\beta }=0$; they are the Maurer-Cartan $1$ forms (\ref{10}) on the Lie
groups $U\left( 1\right) _{\pm }\times SU\left( 2\right) _{L}$ and $U\left(
1\right) _{\pm }\times SU\left( 2\right) _{R}$: 
\begin{equation}
\hat{\Omega}_{L\pm }=g_{\pm }^{-1}\mathbf{d}g_{\pm }=\frac{i}{2a_{\#}}\sigma
_{\alpha }^{\pm }e^{\alpha }\text{,\qquad }\hat{\Omega}_{R\pm }=\bar{g}_{\pm
}^{-1}\mathbf{d}\bar{g}_{\pm }=\frac{i}{2a_{\#}}\bar{\sigma}_{\alpha }^{\pm
}e^{\alpha }\text{.}  \label{16}
\end{equation}

It is important to note that Lagrangian (\ref{1}) contains wedge products of 
\emph{right} and \emph{left} Lie algebra-valued forms: 
\begin{equation}
\begin{array}{c}
\hat{\Omega}_{+}^{L}\wedge \hat{\Omega}_{L}^{+}=\frac{i}{2a_{\#}}\left(
\sigma _{0}e^{0}+\bar{\sigma}_{j}e^{j}\right) \wedge \frac{i}{2a_{\#}}\left(
\sigma _{0}e^{0}+\sigma _{k}e^{k}\right) \\ 
=-\frac{1}{2a_{\#}^{2}}\sigma _{j}\left[ e^{0}\wedge e^{j}+\frac{i}{2}%
\epsilon _{k\ell }^{j}e^{k}\wedge e^{\ell }\right] \text{,}
\end{array}
\label{17}
\end{equation}
for the vacuum spin connections (\ref{16}) on $\mathbb{M}_{\#}$. Note that $%
\Omega ^{L}\wedge \Omega _{L}$ includes both \emph{magnetic} $\left(
e^{k}\wedge e^{\ell }\right) $ and \emph{electric }$\left( e^{0}\wedge
e^{j}\right) $ components. \emph{No electric components would have appeared
without the} $P$\emph{-conjugation in (\ref{17}).}

The wedge product of two left and two right Maurer-Cartan $1$ forms makes
the \emph{Maurer-Cartan }$4$\emph{\ form,} the \emph{scalar} in the Lie
algebra times the volume form: 
\begin{equation}
\Omega ^{4}\equiv \frac{1}{2}Tr\Omega ^{L}\wedge \Omega _{L}\wedge \Omega
^{R}\wedge \Omega _{R}\text{;}  \label{18}
\end{equation}
\begin{equation}
\begin{array}{c}
\text{e.g.\quad }\hat{\Omega}^{4}=\left( \frac{i}{2a_{\#}}\right) ^{4}\frac{1%
}{2}Tr\sigma _{0}e^{0}\wedge \sigma _{1}e^{1}\wedge \bar{\sigma}%
_{2}e^{2}\wedge \bar{\sigma}_{3}e^{3} \\ 
=\frac{i}{16a_{\#}^{4}}\frac{1}{2}Tr\sigma _{0}e^{0}\wedge e^{1}\wedge
e^{2}\wedge e^{3}\equiv \frac{i}{16a_{\#}^{4}}d^{4}V\text{.}
\end{array}
\label{19}
\end{equation}
The $\frac{1}{2}Tr\ $picks out the \emph{scalar} component, $\sigma _{0}$.

By definition, all integrands must be \emph{scalars}, i.e. multiples of the
Clifford unit, $\sigma _{0}$. This is especially clear in \emph{curved space}%
, where the Clifford-algebra frame $\mathbf{\sigma }\left( z\right) $ varies
from point to point. It is a standard calculation to check that $\int \Omega
^{4}$ is invariant with respect to the full \emph{conformal group} of
nonsingular local $E$ transformations, $e^{\alpha \prime }\left( z\right)
=\Lambda _{\beta }^{\alpha }e^{\beta }\left( z\right) $, of the $1$ forms
and their ``internal'' representations (\ref{12}), (A10) on spinor and
spin-vector fields. $\int \Omega ^{4}$ is also $P$, $T$, and $C$ \emph{%
invariant.} Furthermore, \emph{scalar functions,} $f\left( x\right) \Omega
^{4}$, \emph{of the Maurer-Cartan }$4$ \emph{form are the only }$4$\emph{\
forms that can be invariantly integrated!} This is because all natural $4$
forms are scalar multiples of the volume form, (\ref{18}).

Our Lagrangian density, $\Omega ^{4}$, of (\ref{1}) is the invariant measure
on the Einstein group 
\begin{equation}
E=\mathbb{C}\left( U\left( 1\right) \times SU\left( 2\right) \right)
^{4}=GL\left( 2,\mathbb{C}\right) ^{4}.  \label{20}
\end{equation}
Its integral gives the \emph{covering number} $W$ of the group manifold $E$
over spacetime, $\mathbb{M}$. Local extrema of the action integral (\ref{1})
over $\mathbb{M}$ are achieved \cite{clifford}, \cite{xueg} when \emph{all} $%
4$ pairs are $PTC$ symmetric. From (\ref{18}), 
\begin{equation}
\begin{array}{c}
S_{F}\equiv \int_{\mathbb{M}}\mathcal{L}_{g}\overset{PTC}{\longrightarrow }%
\frac{i}{2}\int_{\mathbb{M}}Tr\Omega ^{L}\wedge \Omega _{L}\wedge \Omega
^{R}\wedge \Omega _{R}\equiv \int_{\mathbb{M}}\Omega ^{4} \\ 
=-16\pi ^{3}W\text{.}
\end{array}
\label{21}
\end{equation}
For $\mathbb{M}_{\#}=\mathbb{S}^{1}\times \mathbb{S}^{3}$, 
\begin{equation}
i\int_{\mathbb{S}^{1}\times \mathbb{S}^{3}}\hat{\Omega}^{4}\equiv -16\pi ^{3}%
\text{.}  \label{22}
\end{equation}
Spin frames (\ref{9}) are the fundamental degree-$1$ maps of spin space over 
$\mathbb{M}_{\#}$.

When \emph{singularities }of map (\ref{10}) that assigns spin space
increments to spacetime increments are present, we simply restrict $TS$ to
the \emph{regular region} $\mathbb{M}\equiv \mathbb{M}_{\#}\backslash \cup
D_{J}$ where all 4 spin connections are \emph{defined. }The singular loci $%
D_{J}$ are the supports of \emph{matter fields} in this model.

The global spin connections $\hat{\Omega}$ provide a minimum vacuum energy (%
\ref{22}). But they have another dramatic effect. When wedge products $\hat{%
\Omega}^{4-J}$ multiply local perturbations $\tilde{\Omega}^{J}$, they
effectively quantize their Hodge \emph{dual} fields over Poincar\'{e} \emph{%
dual} cycles $\gamma ^{4-J}$! This happens because products of
Clifford-algebra-valued forms require both their Clifford and Hodge duals to
make the Clifford scalar $\sigma _{0}$ times the volume element (\ref{19}).
This leads to a \emph{residue theorem} below that classifies the topological
obstructions $D_{J}$ to relaxation of the field energy $V_{F}=-S_{F}$, to
the global minimum $16\pi ^{3}$ of (\ref{22}).

\section{Clifford Residues and $de$Rham Cohomology}

The \emph{charges} in nature---electric charge, mass, baryon number,
etc.---are detected by integrating far fields in the regular region, \emph{%
outside} the supports $B_{3}$ of their respective current $3$ forms $\ast
J\left( x\right) $. If the \emph{same} far field could be produced by an
active local spin transformation, $T_{A}\left( x\right) \in E_{A}$, acting
on the vacuum fields around $B_{3}$, then the \emph{same} charge would be
detected within $B_{3}$. What happens inside our singular domains $D_{J}$ is
that the diagonal ($PTC$-symmetric) subalgebra breaks back to the full Lie
algebra of \emph{independent} $L\times R$ spin transformations: 
\begin{equation}
gl\left( 2,\mathbb{C}\right) _{PTC}\overset{D_{J}}{\longrightarrow }gl\left(
2,\mathbb{C}\right) _{L}\oplus gl\left( 2,\mathbb{C}\right) _{R}  \label{23}
\end{equation}
for each of the $J=1$, $2$, $3$, or $4$ chiral pairs that break away from $%
PTC$ conjugacy. We now \emph{remove} open neighborhoods $B_{J}$ containing
each singular locus $D_{J}$, and consider the effect on action integral (\ref
{21}).

Uhlenbeck's theorem and Taubes patching \cite{kron} assure us that we can
replace any vector potential singular inside a domain $D_{J}$ by a \emph{%
regular} connection, and change the action by an integral multiple of $8\pi
^{2}$. We prove an analogous result for spin bundles below.

Suppose $\ell \left( z\right) $ of (\ref{14}) is a section of the (left) $%
gl\left( 2,\mathbb{C}\right) $ spin-frame bundle over the Dirac phase space $%
\mathbb{CM}\subset \mathbb{C}^{4}$, with the singular loci removed \cite
{xueg}. We may write $\ell \left( z\right) $ in polar form as 
\begin{equation}
\begin{array}{c}
\ell \left( z\right) =\ell \left( 0\right) \exp \left( \frac{i}{2}\theta
_{L}^{\alpha }\left( z\right) -\frac{1}{2}\varphi _{L}^{\alpha }\left(
z\right) \right) \sigma _{\alpha } \\ 
\equiv \ell \left( 0\right) \exp \frac{i}{2}\varsigma _{L}^{\alpha }\left(
z\right) \sigma _{\alpha }\text{,}
\end{array}
\label{24}
\end{equation}
just as we may write a complex function $w\left( z\right) $ of one complex
variable as 
\begin{equation*}
\begin{array}{c}
w\left( z\right) =w\left( 0\right) \exp \left( i\theta \left( z\right)
-\varphi \left( z\right) \right) \\ 
\equiv w\left( 0\right) \exp i\varsigma \left( z\right) \text{,}
\end{array}
\end{equation*}
with the phase $\varsigma \left( z\right) $ \emph{complex.}

When phase singularities are present, $\theta $ becomes path-dependent. But $%
\varphi $ does not, provided $w\left( z\right) $ is single-valued. The phase
advance around a $1$-cycle $\gamma $ parametrized by $t$, enclosing $N$
zeroes and $M$ poles of $w$, is the \emph{logarithmic residue} 
\begin{equation}
\int_{\gamma }w^{-1}\left( z\right) dw\left( z\right) \equiv \int_{\gamma
}w^{-1}\left( \frac{dw}{dt}\right) dt=i\int_{\gamma }d\theta \left( z\right)
=i2\pi m\text{,}  \label{25}
\end{equation}
where $m\equiv N-M$. It detects the \emph{winding number of the }$u\left(
1\right) $\emph{\ phase about singularities} by integrating about $1$-cycles
that \emph{lie completely within the regular region.}

The analog for spin bundles $E$ is obtained by integrating $gl\left( 2,%
\mathbb{C}\right) $-valued $m$ forms about cycles $\gamma _{m}$ that \emph{%
lie completely within the regular region} $\mathbb{CM}$. On $\gamma _{m}$, $%
z\equiv \left( z^{0},z^{1},z^{2},z^{3}\right) $ is parameterized by $%
t^{\alpha }$.

We define the integral of a $gl\left( 2,\mathbb{C}\right) $-valued $m$ form $%
\omega ^{m}$ on and $m$-chain $\gamma _{m}$ as the integral of its \emph{%
scalar component,} 
\begin{equation*}
\int_{\gamma _{m}}\omega ^{m}\equiv \frac{1}{2}\int_{\gamma _{m}}Tr\omega
^{m}.
\end{equation*}
Note (A11) that products of \emph{left} and \emph{right} $gl\left( 2,\mathbb{%
C}\right) $-valued forms make the Clifford scalar, $\sigma _{0}$. We call
such products \emph{Clifford-algebra-valued forms.}

We may now state and prove a \emph{residue theorem }for
Clifford-algebra-valued Forms (see \cite{brackx}, \cite{gilbert} for the $%
\gamma _{m}=S_{n-1}$ case):

\begin{description}
\item[\textbf{Theorem 1:}]  The \emph{Clifford residues,} 
\begin{equation}
\begin{array}{c}
\int_{\gamma _{1}}\ell ^{-1}\mathbf{d}\ell =i2\pi m\text{,} \\ 
\int_{\gamma _{3}}r^{-1}\mathbf{d}r\wedge \mathbf{d}r^{-1}r\wedge \ell ^{-1}%
\mathbf{d}\ell =8\pi ^{2}m_{1}m_{2}m_{3}\text{,} \\ 
\text{and\qquad }\int_{\gamma _{4}}\mathbf{d}\ell ^{-1}\ell \wedge r^{-1}%
\mathbf{d}r\wedge \mathbf{d}r^{-1}r\wedge \ell ^{-1}\mathbf{d}\ell =i16\pi
^{3}mm_{1}m_{2}m_{3}
\end{array}
\label{26}
\end{equation}
for bundles of $gl\left( 2,\mathbb{C}\right) $ spin frames over $\mathbb{CM}$
are quantized about $1$-cycles $\gamma _{1}$, $3$-cycles $\gamma _{3}$, and $%
4$-cycles\emph{\ }$\gamma _{4}$. The periods $m$, $m_{1}$, $m_{2}$,\emph{\ }%
and\emph{\ }$m_{3}$ are integers that are invariant under nonsingular\emph{\ 
}$PTC$-symmetric local \emph{deformations,} 
\begin{equation}
\begin{array}{c}
\ell ^{\prime }\left( z\right) =\ell L\left( z\right) , \\ 
r^{\prime }\left( z\right) =R\left( z\right) r=L^{-1}\left( z\right) r\text{,%
}
\end{array}
\label{27}
\end{equation}
provided that $\ell ^{\prime }\left( z\right) $\emph{\ }and\emph{\ }$%
r^{\prime }\left( z\right) $ remain single-valued about $\gamma $.

\item[Proof:]  Using the identity 
\begin{equation}
\exp \left( \frac{i\varsigma }{2}\hat{\varsigma}^{\alpha }\sigma _{\alpha
}\right) =\cos \frac{\varsigma }{2}+i\left( \sin \frac{\varsigma }{2}\right) 
\hat{\varsigma}^{\alpha }\sigma _{\alpha }\text{,}  \label{28}
\end{equation}
we calculate from (\ref{24}) that 
\begin{equation}
\begin{array}{c}
\ell ^{-1}\mathbf{d}\ell \left( z\right) =\left( \frac{i}{2}d\theta ^{\alpha
}\left( z\right) -\frac{1}{2}d\varphi ^{\alpha }\left( z\right) \right)
\sigma _{\alpha } \\ 
\equiv \frac{i}{2}d\varsigma ^{\alpha }\left( z\right) \sigma _{\alpha }%
\text{.}
\end{array}
\label{29}
\end{equation}
Assuming that the column spinors in $\ell \left( z\right) $ are
single-valued on $\gamma _{1}$, the $U\left( 1\right) $ phase advance, $%
\frac{1}{2}\bigtriangleup \theta ^{0}$, about $\gamma _{1}$ must be an
integral multiple of $2\pi $: 
\begin{equation}
\begin{array}{c}
\int_{\gamma _{1}}\ell ^{-1}\mathbf{d}\ell =\int_{\gamma _{1}}\frac{i}{2}%
d\varsigma ^{0}\sigma _{0}=\int_{\gamma _{1}}\left( \frac{i}{2}d\theta ^{0}-%
\frac{1}{2}d\varphi ^{0}\right) \sigma _{0} \\ 
=\frac{i}{2}\left[ \bigtriangleup \theta ^{0}\right] _{\gamma _{1}}-\frac{1}{%
2}\left[ \bigtriangleup \varphi ^{0}\right] _{\gamma _{1}}=i2\pi m\text{.}
\end{array}
\label{30}
\end{equation}
Integral (\ref{30}) is the \emph{period} about a homology $1$-cycle, $\gamma
_{1}\subset H_{1}\left( \mathbb{CM}\right) $, of the nonexact differential $%
1 $ form, 
\begin{equation*}
\Omega \equiv \ell ^{-1}\mathbf{d}\ell \in H^{1}\left( \mathbb{CM}\right) 
\text{,}
\end{equation*}
which belongs to the first deRham cohomology class of $\mathbb{CM}$. Its 
\emph{period,} $m$, is invariant under both homologous deformations, $\gamma
_{1}^{\prime }\in H_{1}\left( \mathbb{M}\right) $, of the cycle, and
nonsingular $E$ perturbations, $\ell ^{\prime }\left( z\right) =\ell L\left(
z\right) $, of the spin frame. If $\gamma _{1}$ is parametrized by time $t$,
integral (\ref{30}) measures the \emph{difference} $\bigtriangleup \theta
^{0}$ of the $U\left( 1\right) $ phase shifts between paths---or the phase
shift along a path that winds around the worldtube $D_{3}\times I$ of a
massive particle. Integral (\ref{30}) then gives the Bohr-Sommerfeld
quantization conditions.

\item  In the \emph{spinfluid regime,} where the dilation/boost flow $%
y^{\alpha }=y^{\alpha }\left( x^{\alpha }\right) $ is a path-dependent
function of $4$ position $x^{\alpha }$, we could choose the $x^{\alpha
}\equiv t^{\alpha }$ as our integration parameters. Alternatively, we could
choose spherical-polar coordinates and parametrize the spatial $3$-ball $%
D_{3}$ by $\left( r,\theta ,\varphi \right) $. $D_{3}$ compactifies to a $3$%
-cycle $\gamma _{3}$ when the perturbed fields must match the vacuum fields
on its boundary.

\item  We get scalar-valued $3$ forms in (\ref{26}) from terms in $\sigma
_{r}e^{r}\wedge \sigma _{\theta }e^{\theta }\wedge \sigma _{\varphi
}e^{\varphi }=i\sigma _{0}e^{r}\wedge e^{\theta }\wedge e^{\varphi }$. For
example, suppose $\gamma _{3}$ contains a radially symmetric $SU\left(
2\right) $ ``hedgehog'' monopole, i.e. a diagonal map from physical space $%
\mathbb{M}_{\#}\backslash 0$ to $\mathbf{\sigma }$-space: 
\begin{equation}
\begin{array}{c}
\ell \left( x\right) =e^{\frac{i}{2}\theta ^{0}\left( x\right) \sigma
_{0}}e^{\frac{i}{2}f\left( r\right) \mathbf{\hat{r}}\cdot \mathbf{\sigma }}
\\ 
r\left( x\right) =e^{\frac{i}{2}\theta ^{0}\left( x\right) \sigma _{0}}e^{%
\frac{i}{2}f\left( r\right) \mathbf{\hat{r}}\cdot \mathbf{\bar{\sigma}}}%
\text{.}
\end{array}
\label{31}
\end{equation}
Then 
\begin{equation*}
\begin{array}{c}
\int_{\gamma _{3}}r^{-1}\mathbf{d}r\wedge \mathbf{d}r^{-1}r\wedge \ell ^{-1}%
\mathbf{d}\ell =-i\int_{I\left( r\right) \times S_{2}\left( \theta ,\varphi
\right) }\mathbf{d}\left[ f\left( r\right) \bar{\sigma}_{r}e^{r}\wedge \bar{%
\sigma}_{\theta }e^{\theta }\wedge \sigma _{\varphi }e^{\varphi }\right] \\ 
=2\pi n\cdot 4\pi m=8\pi ^{2}M\text{,}
\end{array}
\end{equation*}
where $\mathbf{\hat{r}}\cdot \mathbf{\sigma }\equiv \sigma _{r}$, and $f$
has $n$ radial cycles over $I\left( r\right) $.

\item  More generally, the $SU\left( 2\right) $ monopole may have angular
dependence as well. Then 
\begin{equation}
\begin{array}{c}
\int_{\gamma _{3}}r^{-1}\mathbf{d}r\wedge \mathbf{d}r^{-1}r\wedge \ell ^{-1}%
\mathbf{d}\ell =-i\int_{I\left( r\right) \times S_{2}\left( \theta ,\varphi
\right) }\mathbf{d}\left[ f\left( r\right) \mathbf{g}\left( \theta ,\varphi
\right) \cdot \mathbf{\sigma }\right] \\ 
=2\pi n\cdot 4\pi m=8\pi ^{2}M\text{,}
\end{array}
\label{32}
\end{equation}
where $M$ \emph{must be an integer} for $r$ to be single valued on spatial $%
2 $-surfaces $S_{2}\left( \theta ,\varphi \right) =\partial D_{3}$ enclosing
the support of the monopole fields.

\item  Integral (\ref{32}) is quantized because it is the \emph{period} of
the $3$ form $\Omega ^{3}\equiv \frac{1}{2}Tr\Omega ^{\wedge 3}$ over the $3$%
-cycle $\gamma _{3}$. Similarly, the quantization of 
\begin{equation}
\begin{array}{c}
\int_{\gamma ^{4}}\mathbf{d}\ell ^{-1}\ell \wedge r^{-1}\mathbf{d}r\wedge 
\mathbf{d}r^{-1}r\wedge \ell ^{-1}\mathbf{d}\ell \\ 
=\left( i2\pi n\right) \left( 8\pi ^{2}m\right) =i16\pi ^{3}nm\equiv i16^{3}N
\end{array}
\label{33}
\end{equation}
about $4$-cycles $\gamma ^{4}=\gamma ^{1}\times \gamma ^{3}$ follows from
the fact that only terms like 
\begin{equation*}
\sigma _{0}e^{0}\wedge \bar{\sigma}_{1}e^{1}\wedge \bar{\sigma}%
_{2}e^{2}\wedge \sigma _{3}e^{3}=i\sigma _{0}d^{4}V
\end{equation*}
can make a scalar-valued $4$ form. The integer $N$ is the \emph{action }%
contained in $\gamma _{4}$. $\blacksquare $
\end{description}

Heuristically, the reason for this quantization is easy to see: the $u\left(
1\right) \times su\left( 2\right) $ phase gradients of four independent
spinor fields must be stretched over the four orthogonal spacetime
directions in order for $\mathcal{L}_{g}$ of (\ref{1}) to reproduce the $4$%
-volume element. Integrals of these gradients are quantized over the
``vacuum'' $\mathbb{M}\equiv \mathbb{M}_{\#}\backslash \cup D_{J}$ and over
localized $E_{A}$ perturbations, provided that these patch smoothly into the
vacuum phase distribution outside $D_{J}$. Such perturbations may add only
integral units to the action. These integers are invariant under ``small'' $%
E $ transformations (connected to the identity), and may change value by
integer amounts only for the ``large'' $E_{A}$ transformations associated
with introducing another singularity.

On an expanding deformed space we may write our \emph{topological Lagrangian}
$\mathcal{L}_{T}$ in intrinsic coordinates as 
\begin{equation}
\begin{array}{c}
\mathcal{L}_{T}=\frac{i}{2}Tr\Omega ^{L}\wedge \Omega _{R}\wedge \Omega
^{R}\wedge \Omega _{L} \\ 
=\frac{i}{2}\left| -g\right| ^{\frac{1}{2}}Tr\omega ^{L}\wedge \omega
_{R}\wedge \omega ^{R}\wedge \omega _{L}\text{.}
\end{array}
\label{34}
\end{equation}
The $\omega \equiv \omega _{\alpha }E^{\alpha }$ are intrinsic
spin-connection $1$ forms in the coordinate frame of a co-moving observer
and $\left| -g\right| ^{\frac{1}{2}}$ is his $4$-volume expansion factor: 
\begin{equation*}
e^{0}\wedge e^{1}\wedge e^{2}\wedge e^{3}=\left| -g\right| ^{\frac{1}{2}%
}E^{0}\wedge E^{1}\wedge E^{2}\wedge E^{3}\text{.}
\end{equation*}

Noting that $PTC$ symmetry gives invariance of the trace, 
\begin{equation}
R\left( x\right) =L^{-1}(x)\Longrightarrow \mathcal{L}_{T}^{\prime
}=TrL^{-1}(x)\Omega ^{4}L(x)=\mathcal{L}_{T}\text{,}  \label{35}
\end{equation}
we have:

\begin{description}
\item[\textbf{Corollary1:}]  The action 
\begin{equation}
S_{T}\equiv \int_{\mathbb{M}}\mathcal{L}_{T}  \label{36}
\end{equation}
is invariant under the group $E_{P}$ of \emph{passive Einstein
transformations} which connects tetrads and spinors that could represent the 
\emph{same physical state} to observers using different external/internal
coordinate/spin frames. These include the proper $E$ transformations (A15),
cosmic\emph{\ }expansion, \emph{plus} the $P$, $T$, and $C$ reversals which
preserve $PTC$ symmetry: 
\begin{equation}
\begin{array}{c}
\left( q_{\alpha },\xi _{\pm },\chi ^{\pm }\right) \overset{P}{%
\longrightarrow }\left( \bar{q}_{\alpha },\eta _{\pm },\zeta ^{\pm }\right)
\\ 
\left( \xi _{\pm },\eta _{\pm }\right) \overset{T}{\longrightarrow }\left(
\chi ^{\pm },\zeta ^{\pm }\right) \text{,\qquad }\left( \xi _{\pm },\eta
_{\pm }\right) \overset{C}{\longrightarrow }\left( \xi _{\mp },\eta _{\mp
}\right) \text{;}
\end{array}
\label{37}
\end{equation}
\begin{equation}
\left( \xi _{\pm },\eta _{\pm }\right) \overset{PTC}{\longrightarrow }\left(
\zeta ^{\mp },\chi ^{\mp }\right) \text{.}  \label{38}
\end{equation}
\end{description}

Furthermore, since \emph{all} nonzero $4$ forms are proportional to the
volume element, with a local scale factor that may be taken up into $\left|
-g\right| ^{\frac{1}{2}}$ of (\ref{34}), we have:

\begin{description}
\item[\textbf{Corollary 2:}]  Any Lagrangian density on the
multiply-connected space $\mathbb{M}=\mathbb{M}_{\#}\backslash \cup D_{J}$
that is a \emph{natural} (i.e.\emph{. }$E_{P}$-invariant) $4$ form must be
locally\emph{\ }proportional to $\mathcal{L}_{T}$ of (\ref{34}).
\end{description}

\textbf{Corollary 2} \emph{apparently} relieves us mortals of the task of
guessing the ``real'' grand-unified field Lagrangian, and gives us license
to employ $\mathcal{L}_{T}$ as our Lagrangian density outside the singular
loci. The problem is that we mortals apparently cannot experience a $T$
reversed world, and so cannot know $\Omega ^{L}$ and $\Omega ^{R}$ of (\ref
{34}), nor the contribution to $\left| -g\right| ^{\frac{1}{2}}$ from $\dot{y%
}^{0}=\frac{\overset{\cdot }{a}}{a}$, the \emph{rate} of cosmic expansion!
The best we can do is to substitute $\left( \Omega _{L},\Omega _{R}\right) $
for $\left( \Omega ^{L},\Omega ^{R}\right) $ and use the static
approximation 
\begin{equation}
\mathcal{L}_{S}=\frac{i}{2}Tr\Omega _{L}\wedge \Omega _{L}\wedge \Omega
_{R}\wedge \Omega _{R}  \label{39}
\end{equation}
as our Lagrangian density. The action integral 
\begin{equation}
\begin{array}{c}
S_{S}=\frac{i}{2}\int_{\mathbb{M}}Tr\Omega _{0}e^{0}\wedge \Omega
_{1}e^{1}\wedge \Omega _{2}e^{2}\wedge \Omega _{3}e^{3} \\ 
=i\int_{\mathbb{M}^{\prime }}\gamma ^{-4}Tr\omega _{0}E^{0}\wedge \omega
_{1}E^{1}\wedge \omega _{2}E^{2}\wedge \omega _{3}E^{3}
\end{array}
\label{40}
\end{equation}
may be done in either the extrinsic polar $1$ forms $\Omega _{\alpha
}e^{\alpha }$ on $\mathbb{M}$ or in the intrinsic $1$ forms $\omega _{\alpha
}E^{\alpha }$ on our dilated spacetime $\mathbb{M}^{\prime }$. $S_{S}$ is
invariant with respect to \emph{static} dilations $\gamma =\frac{a}{a_{\#}}$
(i.e. \emph{scale} invariant) but cannot pick up $\dot{y}^{0}$, the \emph{%
dilation rate.} $S_{S}$ agrees with the topological action, $S_{T}$, in the $%
T$-symmetric (static) case.

\section{Dual Residues and Charge Quantization}

The global spin connections $\hat{\Omega}$, (\ref{16}) provide the minimum
vacuum energy (\ref{22}). But they have another dramatic effect.

When wedge products $\hat{\Omega}^{4-J}$ multiply perturbations $\tilde{%
\Omega}^{J}$, they effectively quantize these over Poincar\'{e} \emph{dual}
cycles $\ast D^{J}\equiv B_{4-J}$. This happens because products of
``polarized'' Clifford-algebra-valued forms (\ref{16}) require both their
Clifford and Hodge duals to make the Clifford scalar $\sigma _{0}$ times the
volume element---and so contribute to the action.

The vacuum fields can thus be used to ``probe'' inside the singular loci to
produce new invariants---integrals of \emph{Hodge dual fields over
Poincar\'{e}-dual cycles.} These are the \emph{charges.} We prove they are
quantized below.

To account for the polarization of local $J$-fields form $\tilde{\Omega}%
^{J}\left( x\right) $ by the vacuum spin connections $\hat{\Omega}^{4-J}$,
we write each spin connection as a perturbation $\tilde{\Omega}\left(
x\right) $ added onto the global ``vacuum'' distribution $\hat{\Omega}$: 
\begin{equation*}
\Omega =\hat{\Omega}+\tilde{\Omega}\text{.}
\end{equation*}
The perturbed action $\tilde{S}_{T}$ will then have contributions from
products of the $\left( 4-J\right) $ vacuum connections, $J=1$, $2$, $3$, or 
$4$, and $J$ perturbed fields inside each codimension---$J$ singular domain.

If the perturbed fields $\tilde{\Omega}^{J}$ agree with the vacuum fields
outside the singular domain, 
\begin{equation}
\ast D^{J}\equiv B_{4-J}\subset \gamma _{4-J}\text{,}  \label{41}
\end{equation}
then their contributions to the action must be \emph{quantized} by Theorem 1.

The action contributed by each domain $B_{4-J}$ is 
\begin{equation}
\tilde{S}_{T}=\frac{i}{2}\int_{B_{4-J}\times I_{J}}Tr\tilde{\Omega}%
^{J}\left( x\right) \wedge \hat{\Omega}^{4-J}=-16\pi ^{3}m_{J}\text{,}
\label{42}
\end{equation}
where $I_{J}$ is a cycle parametrized by the $J$ variables $\ast x$
orthogonal to $B_{4-J}$.

Now if the perturbed fields $\tilde{\Omega}\left( x\right) $ for $x\in
B_{4-J}$ are \emph{independent} of $\ast x$, the integral over $I_{J}$ may
be \emph{factored out} of (\ref{42}). The result is that the \emph{dual}
current $\left( 4-J\right) $ forms $\ast \tilde{\Omega}^{J}$ become
quantized over their supports $B_{4-J}$. Thus we have

\begin{description}
\item[Theorem 2:]  The \emph{dual residues }$\int_{B_{4-J}}\ast \tilde{\Omega%
}^{J}\equiv Q_{J}$ are quantized, provided the \emph{perturbed} fields agree
(up to a trivial gauge transformation) with the \emph{vacuum} fields outside
a support $B_{4-J}$.

\item[\textbf{Proof:}]  The proof is a calculation which we outline below
for each case.

\begin{description}
\item[$J=1$ Case:]  The $3$ vacuum spin connections create the ``polarized'' 
$3$-volume forms 
\begin{equation}
\hat{\Omega}^{3}=\frac{1}{16a_{\#}^{3}}\sigma _{\alpha }\epsilon _{\;\beta
\gamma \delta }^{\alpha }e^{\beta }\wedge e^{\gamma }\wedge e^{\delta
}\equiv \frac{2}{3}a_{\#}^{3}\sigma _{\alpha }\ast e^{\alpha }\text{,}
\label{43}
\end{equation}
against which the $1$ form field perturbations $\tilde{\Omega}\left(
x\right) \equiv \tilde{\Omega}^{\beta }\left( x\right) \sigma _{\beta }$ are
integrated. Each ``vacuum polarization'' (\ref{43}) picks out its own
internal direction $\sigma _{\alpha }$ in the \emph{trace.}

\item  The resulting contribution, 
\begin{equation}
\tilde{S}_{T}\equiv \frac{2}{3}a_{\#}^{3}\int_{\gamma _{1}\times B_{3}}%
\tilde{\Omega}_{\;\alpha }^{\alpha }e^{\alpha }\wedge \ast e^{\alpha
}=-16\pi ^{3}M\text{,}  \label{44}
\end{equation}
to the action is quantized: $M=\bigtriangleup W$ is an \emph{integer,} if we
require the perturbation to produce an integral change in the covering
number $W$ of internal (spin) space over external spacetime, $\mathbb{M}$.
\end{description}

\item  When the perturbations $\tilde{\Omega}^{J}\left( \mathbf{x}\right) $
are \emph{time independent}, integrating (\ref{44}) over $x^{0}\in \left[
0,4\pi \right] $ gives the \emph{quantized charge} 
\begin{equation}
\int_{B_{3}}\ast J\left( \mathbf{x}\right) \equiv \int_{B_{3}}J_{0}\left( 
\mathbf{x}\right) e^{1}\wedge e^{2}\wedge e^{3}=8\pi ^{2}Q\text{.}
\label{45}
\end{equation}

\item  The charge $Q$ appears as the integral of the current $3$ form $\ast
J\left( x\right) $ \emph{dual} to the $1$ form perturbation (\ref{45}), 
\begin{equation}
\tilde{\Omega}\left( x\right) \equiv J\left( x\right) \equiv J_{\alpha
}\left( x\right) e^{\alpha }\text{,}  \label{46}
\end{equation}
produced when $1$ chiral pair of ``matter spinors'' break away from $PTC$
symmetry inside $B_{3}$. This creates a \emph{bispinor Fermion.} The
quantized charge (\ref{45}) is then the Noether charge under \emph{complex
time translation,} 
\begin{equation}
\begin{array}{c}
\int_{B_{3}}\ast J=\int_{B_{3}}i\left( d\theta ^{0}+id\varphi ^{0}\right)
e^{1}\wedge e^{2}\wedge e^{3} \\ 
=8\pi ^{2}\left( iq-m\right) \text{.}
\end{array}
\label{47}
\end{equation}
Elsewhere (Noether), we identify the real (external) part, $m$, of the
time-translation charge with the \emph{mass} and the imaginary (internal)
part with the \emph{electric charge} of a bispinor Fermion, in the $J=1$
case.

\item  Precisely the \emph{dual} situation arises in the

\begin{description}
\item[$J=3$ Case:]  Here integration against $1$ vacuum connection $\hat{%
\Omega}$ quantizes the integrals of $3$ forms around $1$-cycles, or \emph{%
orbits} $\gamma _{1}$. We identify the quantized integrals of $3$-form
densities in the $J=3$ case as particle energy-momenta $P$. Integrating over 
$\gamma _{1}$ then gives the Bohr-Sommerfeld condition that quantizes the
energy-momentum $1$ forms of \emph{particles} around orbits $\gamma _{1}$.
This is the

\item[$J=4$ Case:]  The quantization of action 
\begin{equation}
\int_{\gamma _{1}}\hat{\Omega}\int_{B_{3}}\tilde{\Omega}^{3}\equiv
\int_{\gamma _{1}}\left( P_{0}e^{0}-P_{j}e^{j}\right) =-16\pi ^{3}N\text{.}
\label{48}
\end{equation}
\end{description}

\item  We examine the most interesting case below, the

\begin{description}
\item[$J=2$ Case:]  Here $2$ forms become quantized over their \emph{dual} $%
2 $ cycles. This gives quantization of \emph{electric} flux---Gauss'
law---after converting $\tilde{\Omega}\wedge \tilde{\Omega}$ to the \emph{%
field} $2$ form $K\equiv \mathbf{d}\tilde{\Omega}+\tilde{\Omega}\wedge 
\tilde{\Omega}$, then integrating by parts. $\blacksquare $
\end{description}
\end{description}

\section{Chern Classes for Bispinor Bundles}

Using expressions (A20) for the spin curvatures, we may rewrite the $T$%
-symmetric ($T_{S}$) part, (\ref{40}), of the action as 
\begin{equation}
\begin{array}{c}
S_{S}=\frac{i}{2}\int_{\mathbb{M}}Tr\Omega _{L}\wedge \Omega _{L}\wedge
\Omega _{R}\wedge \Omega _{L} \\ 
\equiv \frac{i}{2}\int_{\mathbb{M}}Tr\left( K_{L}-\mathbf{d}\Omega
_{L}\right) \wedge \left( K_{R}-\mathbf{d}\Omega _{R}\right) \text{.}
\end{array}
\label{49}
\end{equation}

Using the Bianchi identity $dK=K\wedge \Omega -\Omega \wedge K$, upon
integration by parts (\ref{49}) may be written as 
\begin{equation}
S_{S}=\frac{i}{2}\int_{\mathbb{M}}TrK_{L}\wedge K_{R}+\int_{\mathbb{M}%
}Tr\Omega _{L}\wedge \left( K_{L}+K_{R}\right) \wedge \Omega _{R}+\sum M_{J}%
\text{,}  \label{50}
\end{equation}
where the $M_{J}$ are some Chern-Simons-type integrals about the boundaries
of the singular domains. We showed \cite{clifford}, \cite{xueg}, that the
term in $\left( K_{L}+K_{R}\right) $, the $PT$-symmetric (neutral) part of
the net spin curvature, contains the \emph{Palantini action for gravitation}%
. It vanishes in the $PT$ antisymmetric ($PT_{A}$) case. There $S_{S}$ is
stationarized at 
\begin{equation}
\hat{S}_{A}=\frac{i}{2}\int_{\mathbb{M}}TrK_{L}\wedge K_{R}\equiv -16\pi
^{3}C_{2}  \label{51}
\end{equation}
for the $PT_{A}$, $u\left( 1\right) \times su\left( 2\right) $ phase
perturbations associated with \emph{electroweak} potentials and charges.

$C_{2}$ is the \emph{second Chern number} \cite{eguchi}\emph{\ for the
chiral bispinor bundle} $\psi :\mathbb{M}\longrightarrow L\oplus R$ under
the Clifford-Killing form (\ref{5}), (A11) for the \emph{Minkowsky} metrics.
This requires \emph{wedge products of left and right Lie-algebra-valued 2
forms to make an} $E_{P}$\emph{-invariant} $4$\emph{\ form,} since the
passive Einstein transformations include reciprocal Lorenz \emph{boosts} on
left and right spinors.

The chiral version of the second Chern form is thus the wedge \emph{product}
of the left-and-right $u\left( 1\right) \times su\left( 2\right) $-valued
spin-curvature $2$ forms, 
\begin{equation*}
\begin{array}{c}
K_{L}\equiv \left( K_{L\beta }^{\chi }\right) \sigma _{\chi }e^{\alpha
}\wedge e^{\beta }\text{,} \\ 
K_{R}\equiv \left( K_{R\delta }^{\rho }\right) \bar{\sigma}_{\rho }e^{\gamma
}\wedge e^{\delta }\text{.}
\end{array}
\end{equation*}

The $PT_{A}$ part (\ref{51}) of the action (\ref{49}) is \emph{quantized}
because it is the second Chern number of a bispinor bundle. It resembles the
Yang-Mills action $\int Tr\left( F\wedge \ast F\right) $, with Hodge $\ast $
replaced by $P$ reversal. This resemblance is deeper than it appears, due to
equation (A3).

The spin curvature $2$ forms $K_{L}$ and $K_{R}$ \emph{are} infinitesimal $L$
and $R$ spin transformations; they output infinitesimal $u\left( 1\right)
\times su\left( 2\right) $ holonomy operators about the boundaries of their
input two-cells. They thus naturally decompose into imaginary self-dual
(left) and anti-self-dual (right) parts: 
\begin{equation}
\ast K_{L}=iK_{L}\qquad \ast K_{R}=-iK_{R}\text{,}  \label{52}
\end{equation}
since $\beta \longrightarrow i\beta $ takes us from $SO\left( 4\right) $ in
equations (A2), (A3) to $SO\left( 1,3\right) $.

Furthermore, the unperturbed spin curvatures of canonical connections (\ref
{16}), 
\begin{equation}
\begin{array}{c}
\hat{K}_{L\pm }=-\frac{1}{a_{\#}^{2}}\sigma _{j}\left[ \frac{i}{2}\epsilon
_{\,k\ell }^{j}e^{k}\wedge e^{\ell }\pm e^{0}\wedge e^{j}\right] \\ 
=\left( i\mathbf{B}_{L}+\mathbf{E}_{L}\right) \cdot \mathbf{\sigma } \\ 
\hat{K}_{R\pm }=-\frac{1}{a_{\#}^{2}}\bar{\sigma}_{j}\left[ \frac{i}{2}%
\epsilon _{\,k\ell }^{j}e^{k}\wedge e^{\ell }\mp e^{0}\wedge e^{j}\right] \\ 
=\left( i\mathbf{B}_{R}+\mathbf{E}_{R}\right) \cdot \mathbf{\bar{\sigma}}%
\text{,}
\end{array}
\label{53}
\end{equation}
are \emph{chiral dual:} 
\begin{equation}
\hat{K}_{R\mp }=i\ast \widehat{\bar{K}}_{L\pm }\text{.}  \label{54}
\end{equation}
The ``vacuum fields'' (\ref{53}) are global \emph{dyons,} with equal
electric and magnetic fields distributed over $\mathbb{S}_{3}\left(
a_{\#}\right) $.

$PT_{A}$ perturbations $\kappa $, for which $R^{\mp }=\left( L_{\pm }\right)
^{-1}$, preserve the metric tensor (A11), and therefore the Hodge $\ast $
operator. They thus preserve chiral duality conditions (\ref{54}). For
these, our $U\left( 1\right) \times SU\left( 2\right) $ action on $\mathbb{M}%
_{\#}=\mathbb{S}_{1}\times \mathbb{S}_{3}$ maps to an $\mathbb{R}\times
SU\left( 2\right) $ Yang-Mills action on $\mathbb{R}_{4}$: 
\begin{equation}
\frac{i}{8}\int_{\mathbb{M}_{\#}}Tr\kappa _{L_{\pm }}\wedge \kappa _{R_{\mp
}}\overset{PT_{A}}{\longrightarrow }-\frac{1}{8}\int_{\mathbb{R}%
_{4}}Tr\kappa \wedge \ast \bar{\kappa}\text{.}  \label{55}
\end{equation}
We may thus pull back the t'-Hooft/Jackiw-Noel-Rebbi multi-instanton
solutions \cite{eguchi} on $\mathbb{R}_{4}$ to obtain localized \emph{%
multi-dyon} solutions on $\mathbb{M}_{\#}$. The ``global dyon'' (\ref{53})
centered at $0\in \mathbb{R}^{4}$ combines with a local dyon centered at $%
N\in \mathbb{S}_{3}\left( a_{\#}\right) $, the north pole of our reference
three sphere of radius $\lambda =a_{\#}$, to produce radial spin curvatures
of: 
\begin{equation}
\begin{array}{c}
\kappa _{L\pm }=\left[ \frac{2Ir^{2}}{\left( r^{2}+\lambda ^{2}\right) ^{2}}%
\right] \sigma _{r}\left[ ie^{\theta }\wedge e^{\varphi }\pm e^{0}\wedge
e^{r}\right] \text{,} \\ 
\kappa _{R\pm }=\left[ \frac{2Ir^{2}}{\left( r^{2}+\lambda ^{2}\right) ^{2}}%
\right] \bar{\sigma}_{r}\left[ ie^{\theta }\wedge e^{\varphi }\mp
e^{0}\wedge e^{r}\right] \text{.}
\end{array}
\label{56}
\end{equation}
Here we use spherical-polar coordinates 
\begin{equation}
\begin{array}{c}
\sigma _{r}\equiv \mathbf{\hat{r}}\cdot \mathbf{\sigma }\text{;\qquad }\bar{%
\sigma}_{r}\equiv \mathbf{\hat{r}}\cdot \mathbf{\bar{\sigma}}\text{;} \\ 
e^{0}\wedge e^{r}=dx^{0}\wedge dr\text{,\qquad }e^{\theta }\wedge e^{\varphi
}=r^{2}\sin \theta d\theta \wedge d\varphi
\end{array}
\label{57}
\end{equation}
in both physical space and spin space.

We suggest that the opposite nonAbelian magnetic fields in (\ref{56}) could
bind $L$- and $R$-chirality spinors into charged \emph{bispinor Fermions. }%
Each contributes an action of 
\begin{equation}
\frac{i}{2}\int_{\mathbb{M}_{\#}\backslash 0}\kappa _{L\pm }\wedge \kappa
_{R\pm }=-16\pi ^{3}I^{2}\equiv -16\pi ^{3}C_{2}  \label{58}
\end{equation}
proportional to the \emph{square} of the charge. But there is \emph{also} a
contribution from the interaction of each localized charge with the global
fields (\ref{53})!

We show below how the interaction with the vacuum \emph{magnetic} fields in (%
\ref{53}) quantizes the flux of the \emph{electric} field through any $2$%
-surface that encloses a charge.

\section{Topological Quantization of Electric Flux}

To derive the quantization of electric flux from the quantization (\ref{58})
of action, expand the $PT_{A}$ parts of each spin curvature as the sum of
the vacuum fields $\hat{\kappa}$ of (\ref{53}) and the perturbations $\tilde{%
\kappa}$ due to local sources: 
\begin{equation}
\begin{array}{c}
\hat{\kappa}_{L\pm }+\tilde{\kappa}_{L\pm }\equiv \kappa _{\alpha \beta
}^{\lambda }\sigma _{\lambda }e^{\alpha }\wedge e^{\beta } \\ 
\hat{\kappa}_{R\pm }+\tilde{\kappa}_{R\pm }\equiv \kappa _{\gamma \delta
}^{\rho }\bar{\sigma}_{\rho }e^{\gamma }\wedge e^{\delta }\text{.}
\end{array}
\label{59}
\end{equation}

Substituting ansatz (\ref{59}) into action (\ref{55}), we obtain the cross
terms 
\begin{equation}
\mathcal{S}_{c}=\frac{-1}{2a_{\#}^{2}}\int \left[ \tilde{\kappa}%
_{0j}^{i}+\epsilon _{j}^{k\ell }\tilde{\kappa}_{k\ell }^{j}\right]
_{L}e^{0}\wedge e^{1}\wedge e^{2}\wedge e^{3}+P  \label{60}
\end{equation}
between the local dyon fields and the vacuum fields (since only terms in $%
\sigma _{j}\bar{\sigma}_{j}\ $will contribute to the trace).

In the $PT_{A}$ case, the magnetic fields cancel, but the local \emph{%
electric fields add}: We get a \emph{charged} bispinor particle with a net
``radial hedgehog'' electric field: 
\begin{equation}
\tilde{\kappa}_{Loj}^{\;j}+\tilde{\kappa}_{Roj}^{\;j}=\tilde{\kappa}%
_{oj}^{\;j}\text{.}  \label{61}
\end{equation}
Inserting (\ref{61}) into (\ref{60}), we obtain the local $\wedge $ global
interaction energy density 
\begin{equation}
V_{c}=\frac{1}{2a_{\#}^{2}}\tilde{\kappa}_{0j}^{\;j}e^{0}\wedge e^{1}\wedge
e^{2}\wedge e^{3}\text{.}  \label{62}
\end{equation}

Note that it is the \emph{vacuum magnetic fields} in (\ref{53})---the spin
curvatures of the canonical degree-$1$ maps (\ref{9}) of $SU\left( 2\right) $
over $\mathbb{S}^{3}\left( a_{\#}\right) $---that endows potential energy to
each ``radial hedgehog'' \cite{monto} configuration of\emph{\ electric fields%
} $\tilde{\kappa}_{oj}^{\;j}e^{0}\wedge e^{j}$ floating within it.

For example, suppose that the local electric field $E_{3}\left(
x^{1},x^{2}\right) =\tilde{\kappa}_{03}^{\;3}\left( x^{1},x^{2}\right) $ is
in the $3$ direction in both physical space and spin space, but its
amplitude depends on the coordinates $\left( x^{1},x^{2}\right) $ on a
spatial $2$ surface, $S_{12}$. We may then separate the $PT_{A}$ part of
action (\ref{60}) into the product of integrals over $S_{12}$ and over its
normal coordinates $x^{0}\in \mathbb{S}_{1}\left( a_{\#}\right) $ and $%
x_{3}\in \mathbb{S}_{1}\left( a_{\#}\right) $: 
\begin{equation}
\begin{array}{c}
S_{c}=\frac{-1}{a_{\#}^{2}}\int_{S_{12}}E_{3}\left( x^{1},x^{2}\right)
e^{1}\wedge e^{2}\int_{\mathbb{S}_{1}\left( a_{\#}\right) \times \mathbb{S}%
_{1}\left( a_{\#}\right) }e^{0}\wedge e^{3} \\ 
=-4\pi ^{2}\int_{S_{12}}E_{3}\left( x^{1},x^{2}\right) e^{1}\wedge
e^{2}=-16\pi ^{3}N\text{.}
\end{array}
\label{63}
\end{equation}
$S_{c}$ is quantized over the normal surface $S_{12}$ supporting the
perturbation $E_{3}\left( x_{1},x_{2}\right) $, via conditions (\ref{33}), (%
\ref{51}), (\ref{58}). We have thus derived a version of Gauss' law 
\begin{equation*}
\int_{S_{12}}E_{3}\left( x^{1},x^{2}\right) e^{1}\wedge e^{2}=4\pi N\text{,}
\end{equation*}
where $N$ is an \emph{integer,} because the action (\ref{51}) must change in 
\emph{integral steps} $\bigtriangleup C_{2}$ for each localized ``bubble''
of field patched into the vacuum.

For a sphere $\mathbb{S}^{2}\left( \theta ,\varphi \right) $ of radius $r$
surrounding a charge with \emph{radial} electric field $E_{r}\left( \theta
,\varphi \right) =\tilde{\kappa}_{0r}^{r}\left( \theta ,\varphi \right) $, (%
\ref{60}) gives 
\begin{equation}
\int_{\mathbb{S}^{2}\left( \theta ,\varphi \right) }E_{r}\left( \theta
,\varphi \right) r^{2}\sin \theta d\theta \wedge d\varphi =4\pi N\text{.}
\label{64}
\end{equation}
More generally, (\ref{60}) integrates the spin-space component of the field 
\emph{normal} under spinorization map (\ref{10}) to the spatial area
element: 
\begin{equation}
\int_{S^{2}}\mathbf{E}\cdot \mathbf{dA}=4\pi N\text{,}  \label{65}
\end{equation}
where $S^{2}$ is any $2$-surface enclosing the charge. This is \emph{Gauss'
law}.

Quantization of the normal flux of the electric field over a closed \emph{%
spatial} $2$-surface thus follows directly from the quantization of the
topological action (\ref{51}). It is the \emph{vacuum magnetic fields} in (%
\ref{53}) that convert the integral of the \emph{electric field} $2$ form $%
\kappa _{0j}e^{0}\wedge e^{j}$ into the integral of the \emph{dual} $2$ form 
$\kappa _{0j}e^{\theta }\wedge e^{\varphi }$ over a \emph{spatial} homology
cycle: 
\begin{equation*}
\int_{S^{2}}\ast \tilde{\kappa}=4\pi N\text{,}
\end{equation*}
\emph{Gauss' law.}

After accounting for the action of the homogeneous field (\ref{53}), and the
action (\ref{60}) of localized charges immersed in this field, there is the
remaining contribution of the product of $2$ perturbed fields 
\begin{equation}
S_{K}=\frac{i}{8}\int_{\mathbb{M}}Tr\tilde{\kappa}_{L}\wedge \tilde{\kappa}%
_{R}=-\frac{1}{8}\int_{\mathbb{M}}Tr\tilde{\kappa}_{L}\wedge \ast \widetilde{%
\bar{\kappa}}_{L}\text{.}  \label{66}
\end{equation}
This is the Yang-Mills/Weinberg-Salaam ``field action,'' which is usually
added \emph{by hand} to couple sources to their fields.

Note that the action (\ref{58}) is \emph{quadratic} in the charges, whether
it comes from products (\ref{60}) of the local field interacting with the
global background field, with another charge, or with itself. This offers an
explanation not only of \emph{why} charge$^{2}$ has the units of action, but
an estimate of the unit of charge$^{2}$ divided by the unit of action, i.e.
of the \emph{fine-structure constant,} $\alpha $.

\section{The Fine Structure Constant}

Our spin connections, curvatures, and actions above have all been \emph{%
geometrical} objects in the bundle of \emph{spin frames} over $T^{\ast }%
\mathbb{M}$. No physical units like $e$ or $\hbar $ have explicitly
appeared. The topologically quantized electric charge and action appeared as
``covering numbers'' of internal space over external spacetime $2$- and $4$%
-cycles, respectively. However, since the relative increment $\alpha $ to
the action introduced by adding a single charge in (\ref{63}) is
dimensionless, we might as well compute it in our geometric units.

The action $S_{c}$ of (\ref{60}), due to a unit electric charge immersed in
the global magnetic field, is $16\pi ^{3}$. As in (\ref{66}) \cite{weinberg}%
, we must multiply this by $\frac{1}{4}$ to obtain the Maxwell/Yang Mills 
\emph{field} action produced by a single charge. If we take $\hbar $ as the 
\emph{physical unit} of action, we obtain 
\begin{equation}
\alpha \equiv \frac{e^{2}}{\hbar }=\frac{1^{2}}{4\pi ^{3}}\approx \frac{1}{%
124}  \label{67}
\end{equation}
as the number of units of action produced by adding a unit charge to the
vacuum fields.

The value (\ref{67}) does not agree very well with the observed value, $%
\alpha \approx \frac{1}{137}$. Either the mathematical model for charge
quantization presented here fails to capture ``real world'' physics, or
there is a ``real world'' correction to this model.

But expression (\ref{39}), which we derived for the static ($T$-symmetric)
case, \emph{does} require a correction. When the radius $a\left(
x^{0}\right) $ of our Friedmann universe $\mathbb{S}^{3}\left( a\left(
x^{0}\right) \right) $ is expanding with Minkowsky time $x^{0}$, we need to
include the factor $\dot{y}^{0}$ in the metric tensor. This shows up in $%
\left| g\right| ^{\frac{1}{2}}$ of (\ref{34}), but \emph{not} in our static
scale factor $\gamma ^{-4}$ of (\ref{40}).

This correction arises because our intrinsic tetrads are co-moving with the
Friedmann flow. We thus experience \cite{weinberg} a \emph{Euclidean boost}%
---a tilt of our cotangent frame into the radial ($y^{0}$) direction: 
\begin{equation}
\begin{array}{c}
\left[ 
\begin{array}{c}
dy^{0} \\ 
\left| d\mathbf{x}\right|
\end{array}
\right] ^{\prime }=\left[ 
\begin{array}{cc}
\cos \lambda & \sin \lambda \\ 
-\sin \lambda & \cos \lambda
\end{array}
\right] \left[ 
\begin{array}{c}
dy^{0} \\ 
\left| d\mathbf{x}\right|
\end{array}
\right] \text{,} \\ 
\text{or\quad }\left( dy^{0}+idx\right) ^{\prime }=e^{i\lambda }\left(
dy^{0}+idx\right) \text{,} \\ 
\text{where\quad }\lambda \equiv \tan ^{-1}\left( \frac{a_{\#}}{a}\overset{%
\circ }{a}\right) \equiv \tan ^{-1}\dot{y}^{0}\text{.}
\end{array}
\label{68}
\end{equation}
The Minkowsky-time $1$ form, $e^{0}=dx^{0}=\left| d\mathbf{x}\right| \equiv
dx$, must suffer the same contraction as the spacelike increment to preserve 
$c=1$, and special relativity. Thus our real Minkowsky $4$-volume element $V$
suffers the contraction 
\begin{equation}
\begin{array}{c}
d^{4}V^{\prime }\equiv \func{Re}\left( e^{0}\wedge e^{1}\wedge e^{2}\wedge
e^{3}\right) ^{\prime } \\ 
=\cos 4\lambda \left( e^{0}\wedge e^{1}\wedge e^{2}\wedge e^{3}\right)
\equiv \left( \cos 4\lambda \right) d^{4}V \\ 
\Longrightarrow d^{4}V=\left( \cos 4\lambda \right) ^{-1}d^{4}V^{\prime }%
\text{,}
\end{array}
\label{69}
\end{equation}
when projected to the static reference sphere $\mathbb{S}_{3}\left(
a_{\#}\right) $.

If we, as co-moving observers, could somehow deduce the value $\lambda =\tan
^{-1}\dot{y}^{0}$ of this radial tilt of our cotangent space, we could use (%
\ref{69}) to correct our static approximation (\ref{40}) for cosmic
expansion. But we \emph{can}, because the \emph{spacelike}---or $SU\left(
2\right) $---component of what we observe to be a \emph{lightlike}
translation changes when our spatial hypersurface is tilted with respect to
the invariant null direction! It is precisely this tilt that gave \cite
{weinberg} our correction to the Weinberg angle $\theta _{W}$. This required
a value of $\dot{y}^{0}\approx 0.16$ to match the current best value of $%
28.5^{\circ }$ for $\theta _{W}$. Inserting $\dot{y}^{0}=0.16$ into (\ref{69}%
), we obtain $\left( \cos 4\lambda \right) ^{-1}\simeq 1.11$ as the
correction factor (\ref{69}) for our co-expanding $4$-volume element. This
gives a corrected action of 
\begin{equation}
\left( 1.11\right) 124\simeq 124+13=137  \label{70}
\end{equation}
for a unit charge moving with the Friedmann flow. We can interpret the
additional $13$ units as its ``kinetic energy'' with respect to the
stationery reference sphere $\mathbb{S}^{3}\left( a_{\#}\right) $.

From action (\ref{70}), we obtain 
\begin{equation*}
\alpha ^{\prime }\approx \frac{1}{137}
\end{equation*}
for the fine structure constant, as measured by a co-moving observer. This
is close to the measured value of $\left( 137.037\right) ^{-1}$.

\section{Conclusions and Open Questions}

From a class of Lagrangian densities which reduce to the Maurer-Cartan $4$
form $\Omega ^{4}$ in the $PTC$-symmetric limit, we have derived the
quantization of action and charge. These are simply the covering numbers of
the internal phases in chiral spin bundles over $4$-cycles and $2$-cycles in
the multiply-connected external spacetime $\mathbb{M}\equiv \mathbb{M}%
_{\#}\backslash \cup D_{J}$. It is \emph{electric flux} that is quantized
over spatial surfaces $S_{2}=\partial D_{3}$ surrounding a charge, because
the \emph{vacuum} \emph{magnetic fields} $\hat{\kappa}_{k\ell }$ convert the
electric flux $\tilde{\kappa}_{0j}$ to quantized action. We thus have a
realization of a ``dual topological field theory'' \cite{monto}, \cite
{horowitz}, \cite{tr}, in which Hodge star is replaced by a duality
operation between internal Lie algebras. This is none other than the one
induced by \emph{Clifford} product (A11), in which the tetrads in (A10) 
\emph{are} dyads in some fundamental, global spinor fields.

Thus, the metric tensor (A11) needed to contract two spin-$1$ tensors
(vectors) is itself a spin-$2$ tensor. Any natural $E_{P}$-invariant $4$
form---e.g. a Lagrangian density---\emph{must} be the Clifford-scalar part
of a spin-$4$ tensor, 
\begin{equation}
\mathcal{L}_{g}\in \otimes ^{8}\subset \Lambda ^{4}\text{.}  \label{71}
\end{equation}
We have shown here that the simplest realization (\ref{1}) of such a natural
Lagrangian (\ref{71}) gives quantized actions and charges.

When we add one unit of charge to the vacuum fields (\ref{53}), we increment
the action by $\sim 137$ units, as measured in our intrinsic frame,
co-moving with cosmic expansion. We thus derive a value of $\alpha \sim
\left( 137\right) ^{-1}$ for the fine structure constant.

Is this a numerical coincidence, or is there some relevance to fundamental
physics in the mathematical structure we have developed here? More
basically, do the cosmological background fields $\hat{K}$ really exist, and
do they play a fundamental role in charge quantization? These questions
await further investigation.

\section{Appendix}

Recall \cite{im}, \cite{zahed}, \cite{adkins} that 
\begin{equation*}
\text{chiral }SO\left( 4\right) \equiv \text{Spin }4\sim SU\left( 2\right)
_{L}\times SU\left( 2\right) _{R}/\mathbb{Z}_{2}
\end{equation*}
presents a point $\left( a^{0},\mathbf{a}\right) \in \mathbb{R}^{4}$ as the
``quaternion,'' $q$, and its quaternionic conjugate, $\bar{q}$: 
\begin{equation}
\begin{array}{c}
q=a^{0}\sigma _{0}+i\mathbf{a}\cdot \mathbf{\sigma }\equiv a^{0}\sigma
_{0}+ia^{j}\mathbf{\sigma }_{j}\text{,} \\ 
\bar{q}=a^{0}\sigma _{0}+i\mathbf{a}\cdot \mathbf{\sigma }\equiv a^{0}\sigma
_{0}+ia^{j}\mathbf{\bar{\sigma}}_{j}\text{;} \\ 
j=1,2,3.
\end{array}
\tag{A1}  \label{A1}
\end{equation}
The infinitesimal $so\left( 4\right) $ isometries of $\mathbb{S}_{3}$, 
\begin{equation}
\delta \left[ 
\begin{array}{c}
a^{0} \\ 
\mathbf{a}
\end{array}
\right] =\left[ 
\begin{array}{cc}
0 & -\mathbf{\beta }^{T} \\ 
\mathbf{\beta } & \left[ \mathbf{\alpha }\right]
\end{array}
\right] \left[ 
\begin{array}{c}
a^{0} \\ 
\mathbf{a}
\end{array}
\right] \text{,}  \tag{A2}  \label{A2}
\end{equation}
are presented on the position quaternion, $q$, as 
\begin{equation}
q^{\prime }=Lq\bar{R}=e^{\frac{i}{2}\left( \mathbf{\alpha }+\mathbf{\beta }%
\right) \cdot \mathbf{\sigma }}q\bar{e}^{\frac{i}{2}\left( \mathbf{\alpha }-%
\mathbf{\beta }\right) \cdot \mathbf{\bar{\sigma}}}\text{,}  \tag{A3}
\label{A3}
\end{equation}
with 
\begin{equation*}
\begin{array}{c}
\mathbf{\beta }^{T}\equiv \left( \beta _{1},\beta _{2},\beta _{3}\right) 
\text{,\qquad }\mathbf{\alpha }^{T}\equiv \left( \alpha _{1},\alpha
_{2},\alpha _{3}\right) \text{;} \\ 
\\ 
\left[ \mathbf{\alpha }\right] \equiv \left[ 
\begin{array}{ccc}
0 & \alpha _{3} & -\alpha _{2} \\ 
-\alpha _{3} & 0 & \alpha _{1} \\ 
\alpha _{2} & -\alpha _{1} & 0
\end{array}
\right] \text{.}
\end{array}
\end{equation*}
$\mathbf{\sigma }$ and $\mathbf{\bar{\sigma}}$ generate the \emph{left} and 
\emph{right} Lie algebras---which must be viewed as \emph{completely
independent} in chiral $so\left( 4\right) $, giving $6$ generators in all.
Pure left-spin transformations $\mathbf{\alpha }=\mathbf{\beta }$ correspond
to \emph{self-dual} $2$ forms, under the usual identification of
skew-symmetric matrices $\left[ \mathbf{\alpha }\right] $ with $2$ forms 
\cite{atiyah}. Pure right transformations $\mathbf{\alpha }=-\mathbf{\beta }$
correspond to anti-self-dual $2$ forms.

\emph{Dilations} (e.g. of a Friedmann universe) may be included by adding a
scalar generator $\sigma _{0}$; complexification of which gives an internal $%
U\left( 1\right) $ phase shift. There are $4$ representations, $\exp \left( 
\frac{i}{2}\theta ^{0}\left( z\right) -\frac{1}{2}\varphi ^{0}\left(
z\right) \right) \sigma _{0}$, of translations in complex-time $z^{0}\equiv
x^{0}+iy^{0}$, distinguished by the sign of the internal $u\left( 1\right) $
phase advance with logradius $y^{0}$, $sgn\left( \frac{\partial \theta ^{0}}{%
\partial y^{0}}\right) $, which we identify with the \emph{charge} of the
field, and by the \emph{dilation behavior,} $sgn\left( \frac{\partial
\varphi ^{0}}{\partial x^{0}}\right) $, which distinguishes \emph{leptonic}
(light) from \emph{baryonic} (heavy) spinors. These combine with the two
chiralities to give $\mathbf{8}$ fundamental spinor representations \cite{im}
of the spin isometry group, or \emph{Einstein group,} $E$; the globalization
of the Poincar\'{e} group to a Friedmann universe. These make up the \emph{%
Cartan moving} \emph{spin frames} 
\begin{equation}
\ell ^{\pm },u_{\pm },r^{\pm },v_{\pm }\text{,}  \tag{A4}  \label{A4}
\end{equation}
pairwise. Each spin frame contains two basis spinors with opposite helicity:
the fundamental null modes of the Dirac operators.

To match the standard convention for chiral bispinors on the conformal
compactification \cite{penrose} $\mathbb{M}_{\#}=\mathbb{S}^{1}\times 
\mathbb{S}^{3}\left( a_{\#}\right) $ of Minkowsky space \cite{im}, we write
the leptonic spin frames $\ell ^{\pm }\left( x\right) $ and $r^{\pm }\left(
x\right) $ columnwise as the $GL\left( 2,\mathbb{C}\right) $ matrices 
\begin{equation}
\begin{array}{c}
\begin{array}{c}
\ell ^{\pm }\left( x\right) \equiv \left[ 
\begin{array}{cc}
\ell _{\mathbf{1}}\left( x\right) & \ell _{\mathbf{2}}\left( x\right)
\end{array}
\right] ^{\pm } \\ 
=\sigma _{0}\exp \left( \frac{i}{2a_{\#}}\left( \pm x^{0}\sigma
_{0}+x^{j}\sigma _{j}\right) \right) \equiv \sigma _{0}g_{\pm }\left(
x\right)
\end{array}
\\ 
\begin{array}{c}
r^{\pm }\left( x\right) \equiv \left[ 
\begin{array}{cc}
r_{\mathbf{\dot{1}}}\left( x\right) & r_{\mathbf{\dot{2}}}\left( x\right)
\end{array}
\right] ^{\pm } \\ 
=\bar{\sigma}_{0}\exp \left( \frac{i}{2a_{\#}}\left( \pm x^{0}\bar{\sigma}%
_{0}+x^{j}\bar{\sigma}_{j}\right) \right) \equiv \bar{\sigma}_{0}\bar{g}%
_{\pm }\left( x\right) \text{.}
\end{array}
\end{array}
\tag{A5}  \label{A5}
\end{equation}
We also write the \emph{right} spin frame row-wise as 
\begin{equation}
\begin{array}{c}
\bar{r}\left( x\right) \equiv \left[ 
\begin{array}{c}
r^{\mathbf{\dot{1}}}\left( x\right) \\ 
r^{\mathbf{\dot{2}}}\left( x\right)
\end{array}
\right] \equiv r^{T}\left( x\right) \epsilon ^{T}\text{,} \\ 
\text{where }\epsilon \equiv i\sigma _{2}=\left[ 
\begin{array}{cc}
0 & 1 \\ 
-1 & 0
\end{array}
\right] \text{.}
\end{array}
\tag{A6}  \label{A6}
\end{equation}
The overbar indicates space ($P$) reversal, or \emph{Dirac conjugation}. We
have the Lie-algebra isomorphism: 
\begin{equation}
\bar{\sigma}_{\alpha }\sim \epsilon ^{-1}\left( \sigma _{\alpha }\right)
^{T}\epsilon =\left( \sigma _{0},-\sigma _{1},-\sigma _{2},-\sigma
_{3}\right) \text{.}  \tag{A7}  \label{A7}
\end{equation}

The \emph{moving} spin frames $\ell \left( x\right) $ and $r\left( x\right) $
factor the \emph{moving tetrads}. These are the spin-$1$ tensors 
\begin{equation}
q_{\alpha }\left( x\right) =\sigma _{\alpha \;\mathbf{\dot{B}}}^{\;\mathbf{A}%
}\ell _{\mathbf{A}}\left( x\right) \otimes r^{\mathbf{\dot{B}}}\left(
x\right) \equiv \ell \otimes _{\alpha }\bar{r}\text{:}  \tag{A8}  \label{A8}
\end{equation}
\begin{equation*}
\begin{array}{c}
q_{0}\left( x\right) \equiv \ell _{\mathbf{1}}\left( x\right) \otimes r^{%
\mathbf{\dot{2}}}\left( x\right) -\ell _{\mathbf{2}}\left( x\right) \otimes
r^{\mathbf{\dot{1}}}\left( x\right) \equiv \ell \otimes _{0}\bar{r}\text{,}
\\ 
q_{1}\left( x\right) \equiv \ell _{\mathbf{1}}\left( x\right) \otimes r^{%
\mathbf{\dot{1}}}\left( x\right) +\ell _{\mathbf{2}}\left( x\right) \otimes
r^{\mathbf{\dot{2}}}\left( x\right) \equiv \ell \otimes _{1}\bar{r}\text{,}
\\ 
q_{2}\left( x\right) \equiv i\left( \ell _{\mathbf{1}}\left( x\right)
\otimes r^{\mathbf{\dot{1}}}\left( x\right) -\ell _{\mathbf{2}}\left(
x\right) \otimes r^{\mathbf{\dot{2}}}\left( x\right) \right) \equiv \ell
\otimes _{2}\bar{r}\text{,} \\ 
q_{3}\left( x\right) \equiv \ell _{\mathbf{1}}\left( x\right) \otimes r^{%
\mathbf{\dot{2}}}\left( x\right) +\ell _{\mathbf{2}}\left( x\right) \otimes
r^{\mathbf{\dot{1}}}\left( x\right) \equiv \ell \otimes _{3}\bar{r}\text{;}
\end{array}
\end{equation*}
\begin{equation}
\bar{q}_{\alpha }\left( x\right) =r\left( x\right) \otimes _{\alpha }\bar{%
\ell}\left( x\right) \text{.}  \tag{A9}  \label{A9}
\end{equation}
The \emph{matrix representations} $\mathbf{q}_{\alpha }\left( x\right) $ and 
$\mathbf{\bar{q}}_{\alpha }\left( x\right) $ of the moving tetrads $%
q_{\alpha }\left( x\right) $ and $\bar{q}_{\alpha }\left( x\right) $ have
the matrix elements of the Pauli spin matrices $\sigma _{\alpha }$ and $\bar{%
\sigma}_{\alpha }$ with respect to the \emph{moving} spin frames $\ell
\left( x\right) $ and $r\left( x\right) $.

Under complex $E$ transformations (\ref{12}), the \emph{matrix
representations} of the tetrads with respect to the \emph{original} basis $%
\left( \ell \left( 0\right) ,\bar{r}\left( 0\right) \right) $ are, 
\begin{equation}
\begin{array}{c}
\mathbf{q}_{\alpha }^{\prime }\left( z\right) =L\left( z\right) \mathbf{q}%
_{\alpha }\left( 0\right) \bar{R}\left( z\right) =e^{\frac{i}{2}\zeta
_{L}^{\beta }\left( z\right) \sigma _{\beta }}\sigma _{\alpha }e^{\frac{i}{2}%
\zeta _{R}^{\gamma }\left( z\right) \sigma _{\alpha }} \\ 
\mathbf{\bar{q}}_{\alpha }^{\prime }\left( z\right) =R\left( z\right) 
\mathbf{q}_{\alpha }\left( 0\right) \bar{L}\left( z\right) =e^{\frac{i}{2}%
\zeta _{R}^{\beta }\left( z\right) \sigma _{\beta }}\bar{\sigma}_{\alpha }e^{%
\frac{i}{2}\zeta _{L}^{\gamma }\left( z\right) \sigma _{\gamma }}\text{,}
\end{array}
\tag{A10}  \label{A10}
\end{equation}
where $\zeta ^{\alpha }\left( z\right) \equiv \theta ^{\alpha }\left(
z\right) +i\varphi ^{\alpha }\left( z\right) $. These obey the
anti-commutation relations 
\begin{equation}
\mathbf{\bar{q}}_{\alpha }^{\prime }\mathbf{q}_{\beta }^{\prime }+\mathbf{%
\bar{q}}_{\beta }^{\prime }\mathbf{q}_{\alpha }^{\prime }\equiv \left\{ 
\mathbf{\bar{q}}_{\alpha }^{\prime },\mathbf{q}_{\beta }^{\prime }\right\}
=\left\{ R\mathbf{\bar{q}}_{\alpha }\bar{L},L\mathbf{q}_{\beta }\bar{R}%
\right\} =2g_{\alpha \beta }\sigma _{0}  \tag{A11}  \label{A11}
\end{equation}
of the complexified Clifford algebra of (A10). The metric tensor in (A11) is 
\emph{derived} from the tetrads (A8) and (A9)---which are in turn derived
from the $\mathbf{8}$ fundamental global spinor fields, the dynamical
variables in the theory.

$L$ and $R$ chirality spinors are coupled through the \emph{Dirac operators} 
\begin{equation}
\begin{array}{c}
D\equiv iq^{\alpha }\partial _{\alpha } \\ 
\bar{D}\equiv i\bar{q}^{\alpha }\bar{\partial}_{\alpha }\text{.}
\end{array}
\tag{A12}  \label{A12}
\end{equation}
These are the translation invariant derivations, or \emph{Lie-algebra-valued
vector fields} dual to the Maurer-Cartan forms (\ref{10}).

Covariant derivatives $\nabla _{\alpha }$ automatically appear in the Dirac
operators (A12) by differentiating the Cartan moving spin frames in 
\begin{equation*}
\begin{array}{c}
\xi \equiv \ell _{A}\xi ^{A}\equiv \ell \mathbf{\xi } \\ 
\eta \equiv r_{\dot{A}}\eta ^{\dot{A}}\equiv r\mathbf{\eta }\text{:}
\end{array}
\end{equation*}
\begin{equation*}
\begin{array}{c}
\partial _{\alpha }\xi \equiv \partial _{\alpha }\left( \ell \mathbf{\xi }%
\right) =\ell \partial _{\alpha }\mathbf{\xi }+\left( \partial _{\alpha
}\ell \right) \mathbf{\xi } \\ 
=\ell \left( \partial _{\alpha }+\Omega _{\alpha }\right) \mathbf{\xi }%
\equiv \ell \nabla _{\alpha }\mathbf{\xi }\text{.}
\end{array}
\end{equation*}
The Dirac equations for a bispinor particle are \cite{clifford}, \cite{sachs}%
: 
\begin{equation}
\begin{array}{c}
D\mathbf{\xi }\equiv iq^{\alpha }\left( \partial _{\alpha }+\Omega _{L\alpha
}\right) \mathbf{\xi }=\frac{1}{2a_{\#}}\mathbf{\eta } \\ 
\bar{D}\mathbf{\eta }\equiv i\bar{q}^{\alpha }\left( \bar{\partial}_{\alpha
}+\Omega _{R\alpha }\right) \mathbf{\eta }=\frac{1}{2a_{\#}}\mathbf{\xi }
\end{array}
\tag{A13}  \label{A13}
\end{equation}
in the chiral representation \cite{im}.

\emph{To preserve Einstein covariance of the Dirac equations (A13), we} 
\emph{must write all our matter fields with respect to the same moving spin
frames that factor the spacetime tetrads (A8):} 
\begin{equation}
\begin{array}{c}
\xi _{\pm }\left( x\right) \equiv \ell ^{\pm }\left( x\right) \mathbf{\xi }%
_{\pm }\left( x\right) \equiv \ell ^{\pm }\left( x\right) \left( \mathbf{%
\lambda }_{\pm }+\mathbf{\tilde{\xi}}_{\pm }\left( x\right) \right) \overset{%
g.o.}{\longrightarrow }\tilde{\ell}^{\pm }\left( x\right) \mathbf{\lambda }%
_{\pm } \\ 
\eta _{\pm }\left( x\right) \equiv r^{\pm }\left( x\right) \mathbf{\eta }%
_{\pm }\left( x\right) \equiv r^{\pm }\left( x\right) \left( \mathbf{\rho }%
_{\pm }+\mathbf{\tilde{\eta}}_{\pm }\left( x\right) \right) \overset{g.o.}{%
\longrightarrow }\tilde{r}^{\pm }\left( x\right) \mathbf{\rho }_{\pm }\text{.%
}
\end{array}
\tag{A14}  \label{A14}
\end{equation}
$\mathbf{\lambda }_{\pm }$ and $\mathbf{\rho }_{\pm }$ are the homogeneous
background, or \emph{vacuum,} values of the ``leptonic spinors,'' $\xi _{\pm
}$ and $\eta _{\pm }$. $\mathbf{\tilde{\xi}}_{\pm }$ and $\mathbf{\tilde{\eta%
}}_{\pm }$ are their localized \emph{envelope modulations.} These constitute
electrons $\left( \mathbf{\tilde{\xi}}_{-}\oplus \mathbf{\tilde{\eta}}%
_{-}\right) $, positrons $\left( \mathbf{\tilde{\xi}}_{+}\oplus \mathbf{%
\tilde{\eta}}_{+}\right) $ and neutrinos $\left( \mathbf{\tilde{\xi}}%
_{+}\oplus \mathbf{\tilde{\eta}}_{-}\right) $ in this model \cite{xueg}. The
expressions $\overset{g.o.}{\longrightarrow }$ hold in the \emph{geometrical
optics} (g.o.) regime where no two rays of the same spinor field cross; thus
the phase advance (\ref{12}) along paths is well-defined.

Constant $gl\left( 2,\mathbb{C}\right) $ phase shifts generate the group of
spacetime \emph{isometries,} or \emph{passive Einstein transformations,} $%
E_{P}$. These connect the spin frames that represent the \emph{same state}
to different observers: 
\begin{equation}
\begin{tabular}{cc}
Spatial translations: & $\bigtriangleup \theta _{L}^{j}=\bigtriangleup
\theta _{R}^{j}=\frac{\bigtriangleup x^{j}}{a_{\#}}$ \\ 
Boosts: & $\bigtriangleup \varphi _{L}^{j}=\bigtriangleup \varphi _{R}^{j}=%
\frac{\bigtriangleup y^{j}}{a_{\#}}$ \\ 
Arctime translations: & $\bigtriangleup \theta _{L}^{0}=-\bigtriangleup
\theta _{R}^{0}=\pm \frac{\bigtriangleup x^{0}}{a_{\#}}$ \\ 
Rotations: & $\bigtriangleup \theta _{L}^{j}=-\bigtriangleup \theta _{R}^{j}$%
\end{tabular}
.  \tag{A15}  \label{A15}
\end{equation}

The \emph{conformal dual }spinor to $\xi _{-}$, 
\begin{equation}
\begin{array}{c}
\xi ^{-}\equiv \xi _{-}^{T}\gamma \epsilon \equiv \mathbf{\xi }^{-}\ell _{-}%
\text{,} \\ 
\text{where\quad }\mathbf{\xi }^{-}\equiv \mathbf{\xi }_{-}^{T}\epsilon
\quad \text{and\quad }\ell _{-}\equiv \epsilon ^{-1}\left( \ell ^{-}\right)
^{T}\gamma \epsilon =\left( \ell _{+}\right) ^{-1}\text{,}
\end{array}
\tag{A16}  \label{A16}
\end{equation}
is defined \cite{bade} so that $E$ transformations (A15) along with 
\begin{equation}
\text{Cosmic Expansion:\quad }\bigtriangleup \varphi _{L}^{0}=\bigtriangleup
\varphi _{R}^{0}=\frac{\bigtriangleup y^{0}}{a_{\#}}\text{,}  \tag{A17}
\label{A17}
\end{equation}
are \emph{spin isometries} \cite{xueg}. The $E$ invariance of the $GL\left(
2,\mathbb{C}\right) $ matrix product 
\begin{equation}
\ell _{-}\ell ^{+}=\sigma _{0}  \tag{A18}  \label{A18}
\end{equation}
is what assures that the inner product $\xi ^{+}\xi _{-}$ is $E$ invariant.

The \emph{spin connections} (\ref{11}) may thus be written as 
\begin{equation}
\begin{array}{c}
\Omega _{L_{\pm }}=\tilde{\ell}_{\mp }\mathbf{d}\tilde{\ell}^{\pm }\qquad
\Omega _{R\pm }=\tilde{r}_{\mp }\mathbf{d}\tilde{r}^{\pm }\text{;} \\ 
\Omega ^{L_{\pm }}=\left( \mathbf{d}\tilde{\ell}_{\pm }\right) \tilde{\ell}%
^{\mp } \\ 
\Omega ^{R_{\pm }}=\left( \mathbf{d}\tilde{r}_{\pm }\right) \tilde{r}^{\mp }%
\text{.}
\end{array}
\tag{A19}  \label{A19}
\end{equation}
In \emph{curved} spacetime, where $\mathbf{dd}\neq 0$, these possess \emph{%
spin curvatures} 
\begin{equation}
\begin{array}{c}
\tilde{\ell}_{\mp }\mathbf{dd}\tilde{\ell}^{\pm }=K_{L}^{\pm }=\left( 
\mathbf{d}\tilde{\Omega}_{L}+\tilde{\Omega}_{L}\wedge \tilde{\Omega}%
_{L}\right) ^{\pm } \\ 
\tilde{r}_{\mp }\mathbf{dd}\tilde{r}^{\pm }=K_{R}^{\pm }=\left( \mathbf{d}%
\tilde{\Omega}_{R}+\tilde{\Omega}_{R}\wedge \tilde{\Omega}_{R}\right) ^{\pm }%
\text{.}
\end{array}
\tag{A20}  \label{A20}
\end{equation}

\emph{Effective} spin connections (A19) and curvatures (A20) appear in the $%
g.o.$ regime for each $PTC$\emph{-symmetric} pair of spinor fields in our
Lagrangian $4$ form (\ref{1}). It is products of terms like these that give
the topological forms (\ref{34}), (\ref{51}) for the action in the $PTC$%
-symmetric regime.


\begin{thebibliography}{99}
\bibitem{kron}  S. Donaldson and P. Kronheimer, \emph{The Geometry of
Four-Manifolds}, Clarendon, Oxford (1990).

\bibitem{vdw}  L. Infeld and B.L. Van der Waerden, Sitzber. Preuss. Akad.
Wiss., Physic. Math. K1. 380 (1933).

\bibitem{sachs}  M. Sachs, \emph{General Relativity and Matter,} D. Reidel,
New York (1982).

\bibitem{penrose}  R. Penrose and W. Rindler, \emph{Spinors and Spacetimes,
Volume 2, Spinor and Twistor Methods in Space-Time Geometry}, Cambridge
University Press, Cambridge (1985).

\bibitem{keller1}  J. Keller, ``Spacetime Dual Geometry Theory of Elementary
Particles and Their Interaction Fields,'' \emph{International Journal of
Theoretical Physics} \textbf{23}, 9 (1984).

\bibitem{keller2}  J. Keller, ``Spinors and Multivectors as a Unified Tool
for Spacetime Geometry and for Elementary Particle Physics,'' \emph{%
International Journal of Theoretical Physics} \textbf{30}, 2 (1991).

\bibitem{clifford}  M.S. Cohen, ``Spin Geometry and Grand Unification,'' 
\emph{Advances in Applied Clifford Algebras,} \textbf{11}, 1 (2001).

\bibitem{capp}  M.S. Cohen, ``$\mathbf{8}$ Spinor Grand Unification'', \emph{%
Cosmology and Particle Physics, }CAPP 2000, ed. R. Durrer, J.
Garcia-Bellido, and M. Shaposhnikov (2001).

\bibitem{im}  M.S. Cohen, ``Inertial Mass from Spin Nonlinearity'', \emph{%
International Journal of Modern Physics D} \textbf{7}, 5 (1998).

\bibitem{atiyah}  M.F. Atiyah, N.J. Hitchen, and I.M. Singer, ``Self-Duality
in Four-Dimensional Riemannian Geometry,'' \emph{Proc. Roy. Soc.} (London) 
\textbf{A362} (1978).

\bibitem{xueg}  M.S. Cohen, ``Chiral Unification of Electroweak and
Gravitational Interactions'', \emph{Int. J. Mod. Phys. D } \textbf{8}, 4
(1999).

\bibitem{brackx}  F. Brackx, R. Delanghe, and F. Sommen, \emph{Clifford
Analysis,} Pitman Advanced Publishing Program, Boston (1982).

\bibitem{gilbert}  J. Gilbert and M. Murray, \emph{Clifford Algebras and
Dirac Operators in Harmonic Analysis,} Cambridge University Press, Cambridge
(1991).

\bibitem{eguchi}  T. Eguchi, P.B. Gilkey, and A.J. Hanson, \emph{%
Gravitation, Gauge Theories and Differential Geometry}\textit{,} Physics
Reports, Vol. 66, No. 6, North-Holland Publishing Company, Amsterdam (1980).

\bibitem{monto}  C. Montonen and D. Olive, ``Magnetic Monopoles as Gauge
Particles?'', \emph{Physics Letters }\textbf{72B} 1 (1977).

\bibitem{weinberg}  M.S. Cohen, ``Cosmological Determination of the Weinberg
Angle,'' in \emph{Photon: Old Problems in Light of New Ideas.} ed. V.
Dvoeglazov, NOVA (2000).

\bibitem{horowitz}  G.T. Horowitz, ``Exactly Soluable Diffeomorphism
Invariant Theories,'' \emph{Comm. Math Physics} \textbf{125} (1989).

\bibitem{tr}  M. Temple-Raston, ``Dyons in Topological Field Theories,'' 
\emph{Letters in Mathematical Physics} \textbf{23} (1991).

\bibitem{zahed}  I. Zahed and G.E. Brown, \emph{Physics Reports} \textbf{142}%
, 1 (1986).

\bibitem{adkins}  G.S. Adkins, C.R. Nappi, and E. Witten, \emph{Nuclear
Physics}, \textbf{B228}, 552 (1983).

\bibitem{bade}  W.L. Bade and H. Jehle, ``An Introduction to Spinors,'' 
\emph{Reviews of Modern Physics} \textbf{25}, 3, p.714 (1953).
\end{thebibliography}
\end{document}